\documentclass[amsmath, amssymb,10pt,aps,prb,twocolumn,notitlepage,showpacs,superscriptaddress]{revtex4-1}
\usepackage{graphicx}
\usepackage{calc}
\usepackage{braket}
\usepackage{ulem}   % to strike things out
\usepackage{slashed}
\usepackage{wasysym}
\usepackage{siunitx}
\usepackage{dsfont}
\usepackage{amsthm,amsmath,amsfonts,amssymb,verbatim,color}
\usepackage{graphicx}
\usepackage{subfigure}
\usepackage{bm}
\usepackage{epsfig,slashed}
\usepackage[T1]{fontenc}
\usepackage[colorlinks=true,citecolor=blue,linkcolor=blue,urlcolor=blue]{hyperref}
\usepackage[version=4]{mhchem}
\usepackage{gensymb}
\usepackage{tabularx}

\begin{document}
\title{Broad-band THz emission by Spin-to-Charge Conversion in Topological Material - Ferromagnet Heterostructures}

\author{Xingyue Han}
\affiliation{Department of Physics and Astronomy, University of Pennsylvania, Philadelphia, Pennsylvania 19104, USA}
\author{Xiong Yao}
\affiliation{Department of Physics and Astronomy, Rutgers, The State University of New Jersey, Piscataway, New Jersey 08854, USA}
\author{Tilak Ram Thapaliya}
\affiliation{Department of Physics, University of Miami, Coral Gables, Florida 33146, USA}
\author{Genaro Bierhance}
\affiliation{Department of Physics, Freie Universität Berlin, 14195 Berlin, Germany}
\author{Chihun In}
\affiliation{Department of Physics, Freie Universität Berlin, 14195 Berlin, Germany}
\author{Zhuoliang Ni}
\affiliation{Department of Physics and Astronomy, University of Pennsylvania, Philadelphia, Pennsylvania 19104, USA}
\author{Amilcar Bedoya-Pinto}
\affiliation{NISE Department, Max Planck Institute of Microstructure Physics, Weinberg 2, 06120, Halle, Germany}
\affiliation{Present address: Institute of Molecular Science, University of Valencia, Catedratico Jose Beltrán 2, 46980 Paterna, Spain}
\author{Sunxiang Huang}
\affiliation{Department of Physics, University of Miami, Coral Gables, Florida 33146, USA.}
\author{Claudia Felser}
\affiliation{Max-Planck-Institut fur Chemische Physik fester Stoffe, 01187 Dresden, Germany}
\author{Stuart S. P. Parkin}
\affiliation{NISE Department, Max Planck Institute of Microstructure Physics, Weinberg 2, 06120, Halle, Germany}
\author{Tobias Kampfrath}
\affiliation{Department of Physics, Freie Universität Berlin, 14195 Berlin, Germany}
\author{Seongshik Oh}
\affiliation{Department of Physics and Astronomy, Rutgers, The State University of New Jersey, Piscataway, New Jersey 08854, USA}
\author{Liang Wu}
\email{liangwu@sas.upenn.edu}
\affiliation{Department of Physics and Astronomy, University of Pennsylvania, Philadelphia, Pennsylvania 19104, USA}

\date{\today}

\begin{abstract}

Terahertz spintronic devices combine ultrafast operation with low power consumption, making them strong candidates for next-generation memory technologies. In this study, we use time-domain terahertz emission spectroscopy to investigate spin-to-charge conversion (SCC) in bilayer heterostructures comprising topological insulators (TIs) or Weyl semimetals (WSMs) with ferromagnetic metals (FMs). SCC is studied in TI materials  \ce{Bi2Se3}, Pb-doped \ce{Bi2Se3}, and (Bi$_{1-x}$Sb$_x$)$_2$Te$_3$, and the WSM NbP. Our results reveal that the dependence of SCC on TI thickness varies with interface quality, indicating that thickness dependence alone is not a reliable criterion for distinguishing between inverse spin Hall effect and the inverse Rashba–Edelstein effect mechanisms. We find efficient SCC in TIs depends on both \textit{in-situ} growth to prevent surface oxidation and proper composition. In NbP|FM bilayers, we observe THz emission with efficiency and bandwidth comparable to that of TIs, highlighting the broader potential of topological materials. Finally, broadband spectral measurements demonstrate that both TIs and WSMs can generate THz pulses with frequencies extending up to 8\,THz. These findings underscore the promise of topological materials as efficient platforms for ultrafast, broadband spintronic applications.

\end{abstract}

\maketitle

\textbf{Introduction}
Spintronic devices have attracted considerable attention due to their potential for low power consumption and high-speed operation \cite{zutic2004RMP, dieny2020natelec}. Central challenges in this field include the generation, manipulation, and detection of spin currents. Among these, the interconversion between spin and charge currents plays a critical role in both the creation and readout of spin signals \cite{han2018npjquan, sanchez2013NatCom}. Spin–charge interconversion is typically realized via mechanisms such as the spin Hall effect (SHE), which occurs in the bulk of materials with strong spin–orbit coupling (SOC) \cite{sinova2015RMP}, and the Rashba–Edelstein effect (REE), which emerges at interfaces with broken inversion symmetry \cite{EDELSTEIN1990ssc}, along with their reciprocal processes—the inverse spin Hall effect (ISHE) and the inverse Rashba–Edelstein effect (IREE) \cite{seifert2016NatPho, sanchez2013NatCom}.

The discovery of topological insulators (TIs) has introduced a new class of materials with high spin–charge conversion (SCC) efficiency due to strong SOC \cite{zhu2015SciRep, wang2018AdvMat}. Multiple mechanisms have been proposed to account for SCC in these systems. On the one hand, the topological surface states (TSS), characterized by helical spin textures and spin–momentum locking, can contribute to SCC \cite{wang2017natcomm, noyan2025nanoletters, shiomi2014prl, mellnik2014nature, wang2016prl, lee2015prb}. On the other hand, topologically trivial bulk states and surface two-dimensional electron gas (2DEG) have also been shown to contribute to SCC \cite{wang2017natcomm, deorani2014prb, jamali2015nl}. As a result, the relative contributions of surface (REE) and bulk (ISHE) effects remain an active subject of debate. A unifying perspective grounded in bulk–surface correspondence has even suggested that the roles of surface and bulk states cannot be separated and should instead be regarded as equivalent \cite{wang2019prr}.

Experimental investigations of SCC in TI typically employ heterostructures, in which a magnetic layer is placed adjacent to the TI layer. The role of the magnetic layer depends on the measurement technique: it acts as a spin current source in THz emission experiments \cite{wang2018AdvMat, sahu2023acsami} and spin pumping measurements \cite{jamali2015nl, shi2025prb, deorani2014prb, sahu2023acsami, shiomi2014prl, wang2016prl}, or as a spin-torque detector in spin-torque ferromagnetic resonance  measurements \cite{noyan2025nanoletters, shi2025prb, wang2017natcomm, mellnik2014nature}. Among these techniques, THz emission spectroscopy offers a unique advantage as a contact-free, ultrafast probe of SCC \cite{seifert2016NatPho, ni2021NatCom}. In this approach, an ultrashort laser pulse excites the ferromagnetic metal (FM) layer, generating a spin voltage and hence a spin-polarized current, which is subsequently converted into a transverse charge current in the adjacent TI layer \cite{rouzegar2022prb, beaurepaire2004apl}. The resulting sub-picosecond charge current emits a transient electromagnetic pulse in the THz frequency range. The amplitude of this THz signal provides a direct measure of SCC efficiency. Previous THz emission studies have reported that a \ce{Bi2Se3}(10\,QL)|Co(3\,nm) bilayer exhibits SCC efficiency comparable to that of the widely used Pt(6\,nm)|Co(3\,nm) system \cite{wang2018AdvMat}, and have attributed this performance to the contribution of the TSS. However, the influence of the FM layer on the TSS has often been neglected. Theoretical studies suggest that the presence of a metallic overlayer (e.g., Co or Ni) induces charge transfer at the interface, which shifts the TI Fermi level into the conduction band, pushing the TSS away from the Fermi level and may even destroy the helical spin structure \cite{zhang2016prb, culcer2010prb, eremeev2013prb, luo2013prb, hsu2017prb, spataru2014prb}. These effects introduce ambiguity in determining whether SCC originates primarily from surface or bulk states in TI|FM heterostructures.

In addition to TIs, another class of topological materials—Weyl semimetals (WSMs)—offers exciting potential for SCC, but remains comparatively less explored. Like TIs, WSMs exhibit band inversion and strong SOC \cite{yan2017ar, wan2011prb, armitageRMP2018}. Their topologically nontrivial band structures feature Weyl points that act as monopoles of Berry curvature in momentum space. This gives rise to a large intrinsic spin Hall effect \cite{sun2016prl, han2022prb, mannaPRX2018, liu2018giant}, suggesting that WSMs could serve as efficient platforms for SCC and broadband THz emission.

\begin{figure}
\centering
\includegraphics[width=0.5\textwidth]{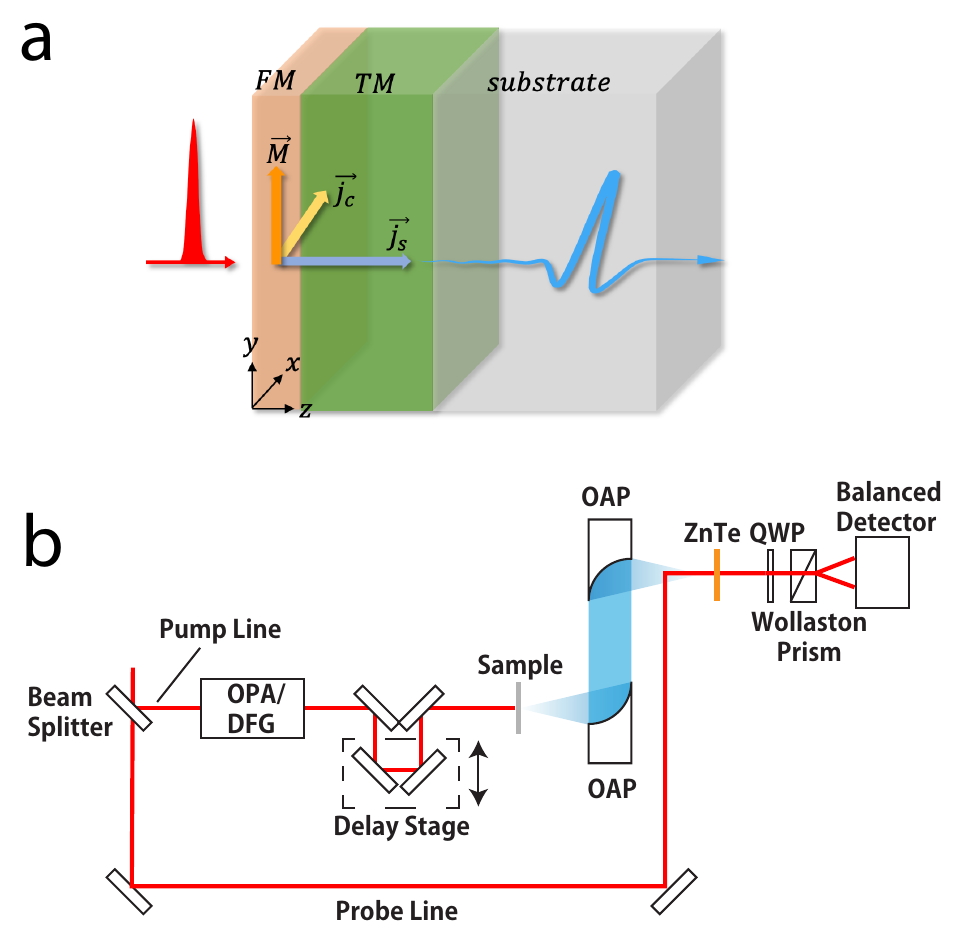}
\caption{
 \textbf{a.} Bilayer structure in ferromagnet (FM)| topological material (TM) films. \textbf{b.} Schematic of the THz emission setup. }
\label{Fig0}
\end{figure}

In this work, we employ all-optical time-domain THz emission spectroscopy \cite{ni2021NatCom} to investigate SCC in TI|FM and WSM|FM heterostructures on a sub-picosecond timescale (Fig.~\ref{Fig0}\textbf{a}). We choose amorphous CoFeB as the FM layer due to its high spin polarization \cite{huang2008spin} and resistance to oxidation \cite{jen2006jap}. We begin by investigating \ce{Bi2Se3}|CoFeB bilayers, identifying two distinct contributions to the THz signal: one from SCC and another from a shift current. By isolating these components, we find that they exhibit different dependencies on TI thickness. Importantly, we show that \textit{in-situ} growth is crucial for achieving strong THz emission. We further study a series of (Bi$_{1-x}$Sb$_x$)$_2$Te$_3$|CoFeB heterostructures with different compositions. The THz emission efficiency reaches a minimum when $x=0.9$, where the chemical potential is close to the charge neutrality point in pristine (Bi,Sb)$_2$Te$_3$ single layer. A similar suppression is observed in Pb-doped \ce{Bi2Se3}|CoFeB samples where bare Pb-doped \ce{Bi2Se3} has insulating bulk. Finally, we demonstrate that a WSM|FM bilayer composed of NbP and permalloy (Py) can also produce strong broadband THz emission. Our spectral analysis confirms that both TI and WSM heterostructures emit radiation up to 8\,THz. These results highlight the potential of topological materials for next-generation ultrafast spintronic applications.

\textbf{Sample preparation and experiment setup}

The TI bilayer sample growth proceeds in two stages. First, \ce{Bi2Se3}, (Bi,Sb)$_2$Te$_3$, and Pb-doped \ce{Bi2Se3} are synthesized on $10 \times 10$ $\text{mm}^2$ $c$-plane (0001) sapphire (\ce{Al2O3}) substrates via molecular beam epitaxy (MBE). The films are then transferred from the MBE chamber to a sputtering system for the second stage, where a 4 nm amorphous CoFeB layer is deposited on top of the TI film. During the transfer, some TI films are left unprotected and thus exposed to air; we refer to these as \textit{ex-situ} grown samples. Other TI films are protected by a Se (for \ce{Bi2Se3} and Pb-doped \ce{Bi2Se3}) or Te (for (Bi,Sb)$_2$Te$_3$) capping layer. After transfer, the capping layer is removed by heating in the sputtering system, yielding the \textit{in-situ} grown samples. The WSM bilayer sample NbP(5 nm)|Py(6 nm) is grown \textit{in-situ}. The single-crystalline NbP layer is grown on a MgO (100) substrate by MBE \cite{bedoyapinto2020acsnano}.

\begin{figure*}
\centering
\includegraphics[width=0.8\textwidth]{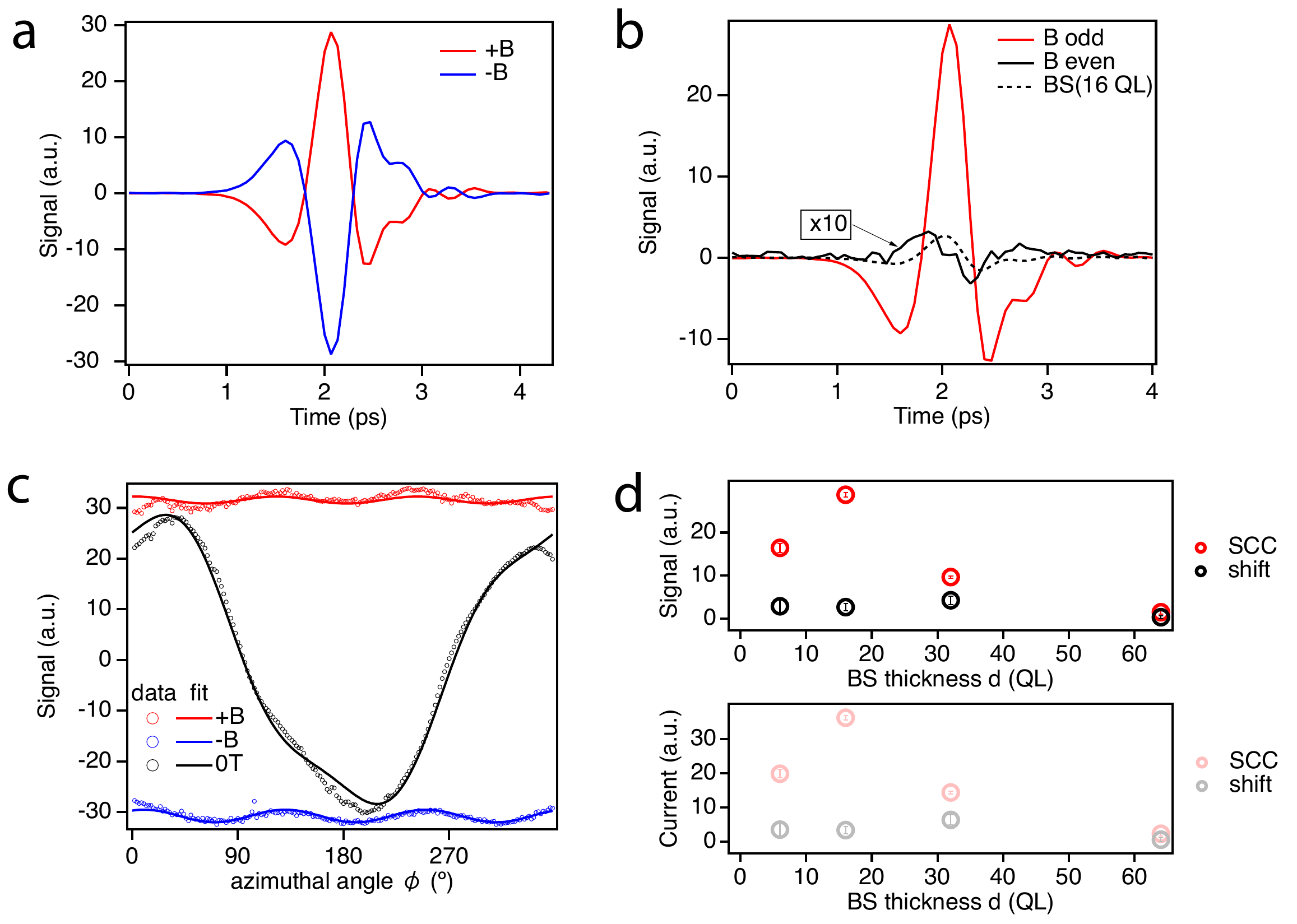}
\caption{
\textbf{a.} THz emission from an \textit{in situ} grown \ce{Bi2Se3}(16 QL)|CoFeB(4 nm) sample under opposite magnetic field directions. 
\textbf{b.} Decomposition of the THz emission into field-symmetric (even) and field-antisymmetric (odd) components. The B-even curve is magnified by a factor of 10 for clarity. The dashed line shows the THz emission from a single-layer \ce{Bi2Se3}(16 QL) film.
\textbf{c.} Azimuthal angle dependence of the peak THz signal with the external field applied in $+\mathbf{B}$ and $-\mathbf{B}$ directions, as well as in zero-field (0 T). The circles represent experimental data, while the curves are fits using Eqs.~\ref{Eq3} and \ref{Eq4}. 
\textbf{d.} SCC and shift-current contributions to the THz signal (top) and extracted charge current (bottom) in \ce{Bi2Se3}($d$\,QL)|CoFeB(4\,nm) samples with varying TI thicknesses ($d = 6$, 16, 32, 64\,QL). Data points are averaged over two samples grown under identical conditions; error bars indicate the standard deviation. Pump photon energy: 0.8\,eV.
}
\label{Fig1}
\end{figure*}

FIG.\ref{Fig0}\textbf{b} illustrates the THz emission spectroscopy setup used for all measurements except for FIG.\ref{Fig4}\textbf{b}. An ultrafast laser (center photon energy of 1.55 eV, pulse duration of 35 fs, repetition rate of 1 kHz) is split into pump and probe beams. On the pump side, an optical parametric amplifier (OPA) converts the photon energy to 0.47--1.55 eV, and subsequent difference-frequency generation (DFG) yields photon energies in the range of 0.20--0.48 eV. The pump beam is focused onto the sample at normal incidence with a spot size of approximately 1 mm and an average power of 7 mW. Its polarization lies along the $\hat{x}$ direction, while a static magnetic field of 0.02 T is applied along $\hat{y}$ using a permanent magnet. The emitted THz wave is collected by one off-axis parabolic mirror (OAP) and then focused by a second OAP onto a electro-optical (EO) crystal. We use 1 mm ZnTe as the EO crystal. The probe beam co-propagates with the THz field into the EO crystal, where the THz electric field induces birefringence, thereby changing the polarization of probe beam. This polarization change is detected in the birefringence measurement using a quarter-wave plate (QWP), a Wollaston prism, and a balanced photodetector. All measurements are performed at room temperature in a dry-air environment with relative humidity maintained below 3\% to minimize water absorption. A THz wire-grid polarizer (not shown) is placed in the THz path to select the THz electric field component along the $\hat{x}$ direction. By scanning a motorized delay stage in the pump line, we record the time-domain profile of the THz electric field. Another similar setup using 250 $\mu$m GaP is used for broadband THz detection for FIG.\ref{Fig4}\textbf{b}. More details about the broadband setup are included in the Supplementary Material. We also measure the THz transmission of the films and calculate the THz impedance to extract the charge current. More details about the THz transmission setup can be found in our previous paper \cite{han2022prb, han2024nanolet, stensberg2023prl}.\\

\textbf{Results}

We first present the THz emission results obtained from the \textit{in-situ} grown \ce{Bi2Se3}(16 QL)|CoFeB(4 nm) sample using 0.8 eV pump. By applying an external magnetic field in the $\pm \hat{y}$ direction using a permanent magnet, we observe a sign change in the emitted THz signal, confirming its magnetic origin (Fig.~\ref{Fig1}\textbf{a}). We decompose the measured signals into field-symmetric (B-even) and field-antisymmetric (B-odd) components (Fig.~\ref{Fig1}\textbf{b}). The field-antisymmetric term is half of the difference between the two traces in Fig.~\ref{Fig1}\textbf{a} while the field-symmetric term is half of the sum of the two time traces.   The B-odd contribution is two orders of magnitude larger than the B-even one, indicating that the magnetism-driven THz emission dominates. We attribute the B-odd component to SCC, and the B-even component to the shift current from the bare \ce{Bi2Se3} surface \cite{braun2016ulNatCom, stensberg2024prb}. For comparison, we also show the THz emission from a single-layer \ce{Bi2Se3}(16 QL) film in Fig.~\ref{Fig1}\textbf{b}. Since SCC is absent in the single-layer sample, the observed signal arises purely from the surface shift current. Its amplitude is roughly an order of magnitude smaller than the SCC contribution in the bilayer and exhibits a three-fold symmetry imposed by the crystal lattice. This confirms that the 4 nm CoFeB layer serves effectively as the spin current source, even in the absence of a protective capping layer \cite{jen2006magnetic}.

The coexistence of SCC and shift current also explains the sample azimuthal angle $\phi$ dependence shown in Fig.~\ref{Fig1}\textbf{c}. Under the external field $\vec{B}=B\hat{y}$, the spin current $\vec{j_s}$ flows along $\hat{z}$ with the spin polarization $\vec{\sigma}=\sigma\hat{y}$ (see FIG.\ref{Fig0}\textbf{a}). The SCC-induced THz emission direction ($\propto \vec{j_s} \times \vec{\sigma} \propto \hat{z} \times \hat{y} \propto \hat{x}$) is then fixed along $\hat{x}$, and thus exhibits no azimuthal dependence \cite{seifert2016NatPho}. Meanwhile, the shift current originating from the \ce{Bi2Se3} surface follows a three-fold symmetry that depends on the crystal orientation \cite{braun2016ulNatCom, stensberg2024prb}. Consequently, under $\pm B$ (red and blue circles in Fig.~\ref{Fig1}\textbf{c}), the peak signal $S_{\pm B}(\phi)$ can be described by 
\begin{equation}
    S_{\pm B}(\phi) = \pm S_{\mathrm{SCC}} + S_{\mathrm{shift}}\sin\bigl[3(\phi - \phi_0)\bigr],
\label{Eq3}
\end{equation}
where $S_{\mathrm{SCC}}$ is the SCC contribution whose sign changes with $B$ polarity, $S_{\mathrm{shift}}$ is the shift current contribution, and $\phi_0$ is an offset between the crystal axes and the laboratory frame. When the external field is removed, the magnetization $\vec{M}$ and spin current polarization $\vec{\sigma}$ rotates with the sample azimuthal angle $\phi$, $\vec{\sigma}=\sigma [\cos{(\phi-\phi_0')}\hat{x} + \sin{(\phi-\phi_0')}\hat{y}]$. Following the same analysis above, at 0 T (black circles in Fig.~\ref{Fig1}\textbf{c}), the peak signal $S_{0T}(\phi)$ can be described as
\begin{equation}
    S_{0T}(\phi) = S_{\mathrm{SCC}}\sin\bigl(\phi - \phi_0'\bigr) + S_{\mathrm{shift}}\sin\bigl[3(\phi - \phi_0)\bigr],
\label{Eq4}
\end{equation}
where $\phi_0'$ is an additional offset capturing the remanent magnetization direction. The fitting results are plotted with solid lines in FIG.\ref{Fig1}\textbf{c} which agrees with the data.

We apply this analysis to \ce{Bi2Se3}($d$\,QL)|CoFeB(4\,nm) samples with varying TI thicknesses, as shown in Fig.~\ref{Fig1}\textbf{d}. Here, the \textit{signal} refers to the measured emitted terahertz electric field, while the \textit{current} denotes the extracted transverse charge sheet current within the heterostructure, taking into account its impedance \cite{seifert2016NatPho}. As $d$ increases from 6\,QL to 64\,QL, the SCC component exhibits a nonmonotonic trend, peaking around $d = 16$\,QL. In contrast, the shift current shows no clear thickness dependence, consistent with its origin from inversion symmetry breaking at the surface. For thin samples (6\,QL and 16\,QL), SCC clearly dominates the THz emission; at $d = 32$\,QL, SCC remains approximately twice as large as the shift current contribution; and at $d = 64$\,QL, the two components become comparable. A similar trend is observed for the extracted sheet current, reinforcing this thickness-dependent behavior. Our results differ from previous observations in \ce{Bi2Se3}($d$\,QL)|Co(4\,nm) bilayers, where the THz signal decreases monotonically beyond $d = 6$\,QL \cite{wang2018AdvMat}. The weak thickness dependence of the shift current, in contrast to the pronounced nonmonotonicity of SCC, suggests that the SCC in our samples is not dominated by a surface-driven mechanism.

\begin{figure}
\centering
\includegraphics[width=0.45\textwidth]{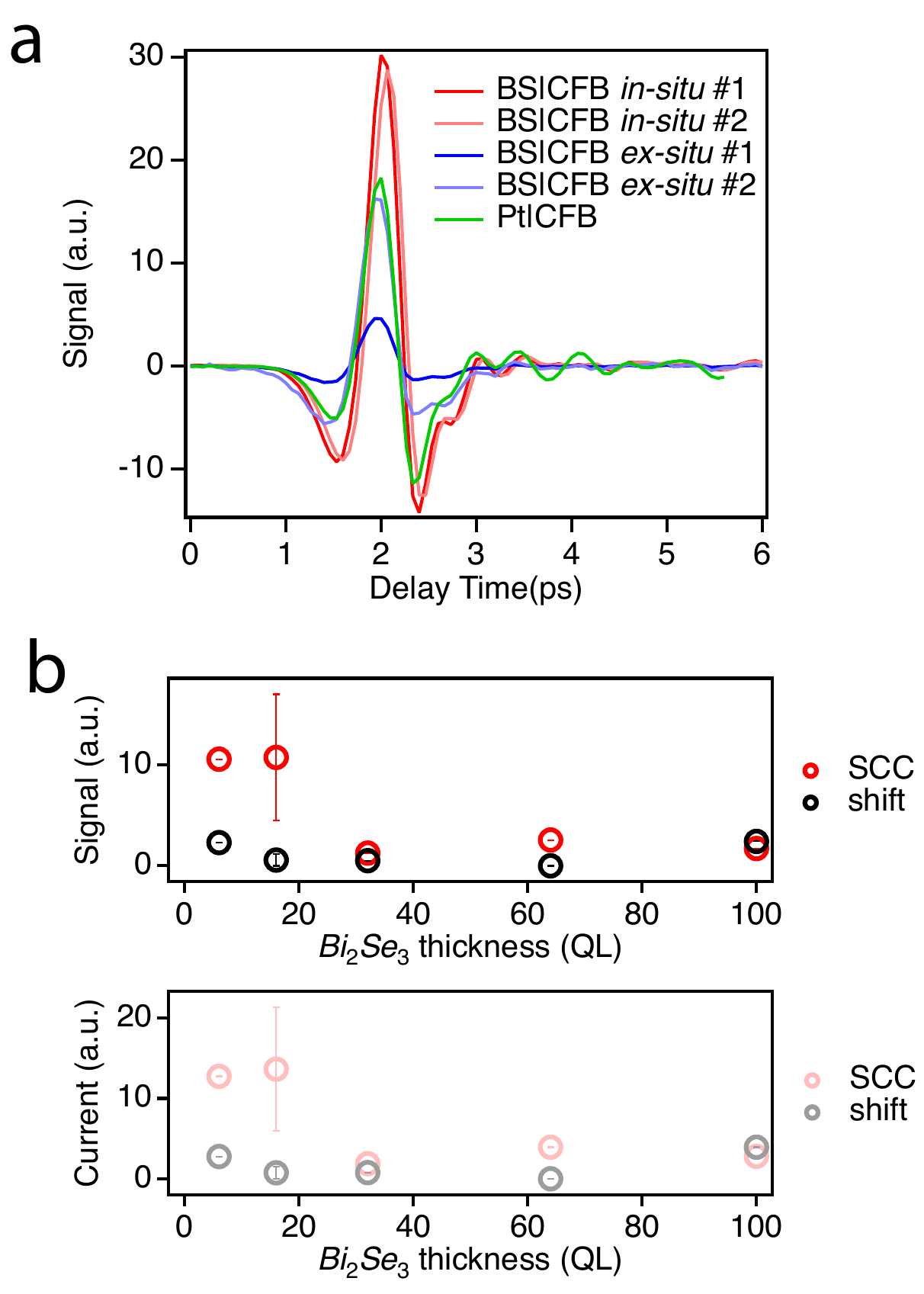}
\caption{
\textbf{a.} Time-domain THz waveforms for two \textit{in situ} and two \textit{ex situ} \ce{Bi2Se3}(16 QL)|CoFeB(4 nm) films and one Pt(16 nm)|CoFeB(4 nm) reference. 
\textbf{b.} SCC and shift-current contributions to the THz signal (top) and extracted current (bottom), obtained from the azimuthal-angle dependence in \textit{ex-situ} grown \ce{Bi2Se3}($d$\,QL)|CoFeB(4\,nm) samples with varying TI thickness $d$. The data point at $d = 16$\,QL is averaged over the two samples shown in \textbf{a}; the corresponding error bar indicates their standard deviation. Other thicknesses are represented by a single sample and thus do not include error bars. Pump photon energy: 0.8\,eV.
}
\label{Fig2}
\end{figure}

\begin{figure*}
\centering
\includegraphics[width=0.8\textwidth]{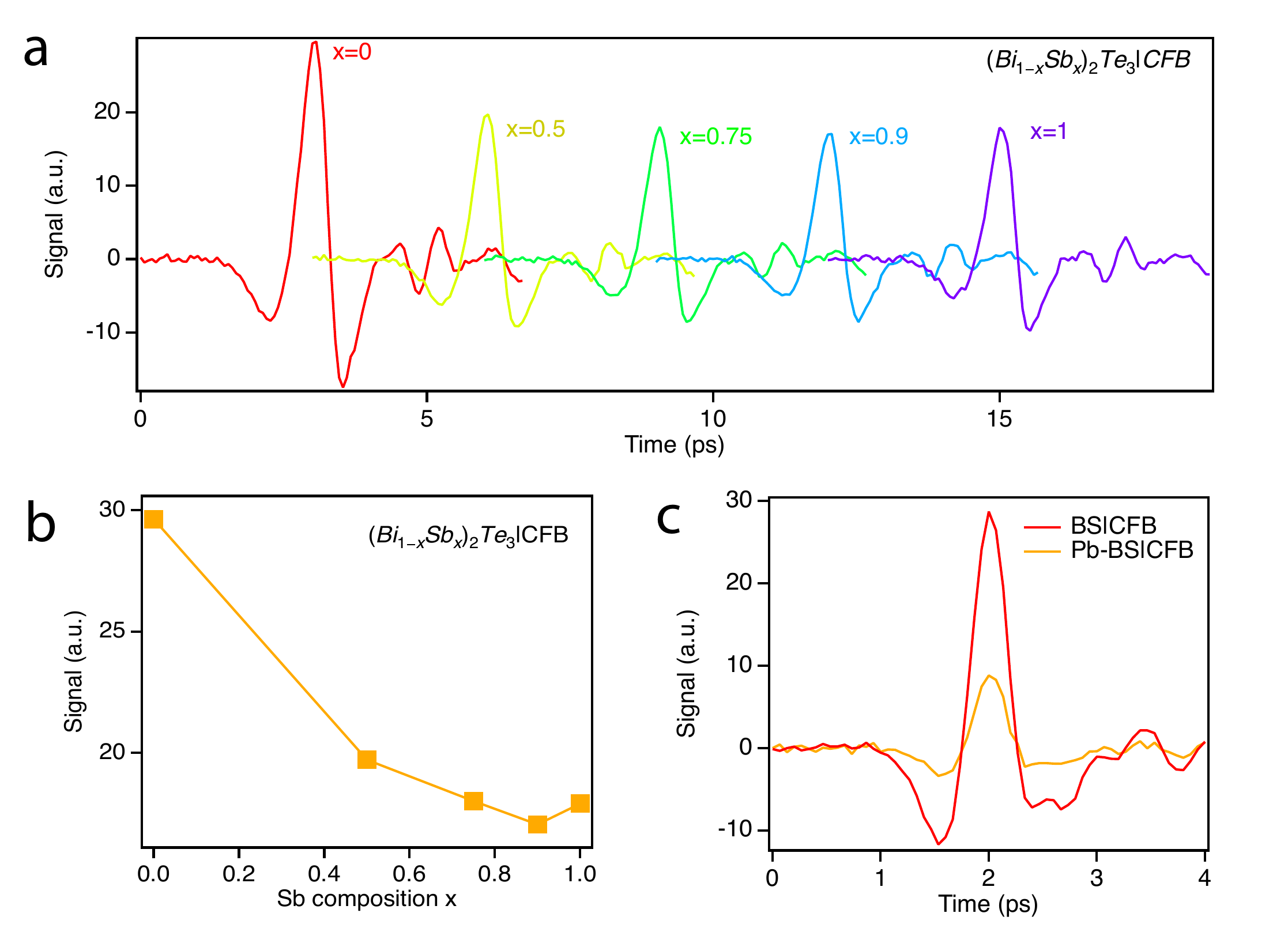}
\caption{
\textbf{a.} Time-domain THz waveforms for \ce{(Bi_{1-x}Sb_x)_2Te_3}(16 QL)|CoFeB(4 nm). 
\textbf{b.} Sb composition ($x$) dependence of the THz emission from \ce{(Bi_{1-x}Sb_x)_2Te_3}(16 QL)|CoFeB(4 nm).
\textbf{c.} Time-domain THz waveforms for \ce{Bi2Se3}(16 QL)|CoFeB(4 nm) and Pb-\ce{Bi2Se3}(16 QL)|CoFeB(4 nm). Pump photon energy: 1.55 eV.}
\label{Fig3}
\end{figure*}

To investigate the impact of surface condition on THz emission, we prepared \textit{ex-situ} grown \ce{Bi2Se3}(16\,QL)|CoFeB(4\,nm) films for comparison. The key difference between \textit{ex-situ} and \textit{in-situ} grown samples lies in whether the TI surface was exposed to air prior to capping, and thus potentially oxidized. While \textit{in-situ} grown samples consistently exhibit high THz emission efficiency, their \textit{ex-situ} counterparts display significantly reduced signal amplitudes, as illustrated in Fig.~\ref{Fig2}\textbf{a} for representative \ce{Bi2Se3}(16\,QL)|CoFeB(4\,nm) samples. Notably, the two \textit{ex-situ} samples were grown in separate batches and exhibit large variations in signal strength, likely due to different durations of air exposure. Measurements of THz transmission and impedance confirm that both \textit{ex-situ} grown samples are comparable in out-coupling current to THz emission (see Supplementary Material, Fig.~S1), ruling out these factors as the cause of the observed discrepancy in emission.  Although topological surface states are theoretically protected from oxidation \cite{wang2012prl, yashina2013acsnano, yang2020jpcc, green2016jvst}, the suppressed emission observed in \textit{ex-situ} grown samples may result from reduced spin-current injection efficiency across the oxidized FM/TI interface. As shown in Fig.~\ref{Fig2}\textbf{b}, this variability introduces significant uncertainty and makes it difficult to extract consistent thickness-dependent trends \cite{wang2018AdvMat}. In conclusion, our results demonstrate that the observed thickness dependence of THz emission is highly sensitive to sample growth conditions. This sensitivity compromises the reliability of using thickness-dependent measurements alone to distinguish between ISHE and IREE mechanisms in \ce{Bi2Se3}|CoFeB heterostructures \cite{wang2018AdvMat}. %\red{Comment out that the results are similar under 0.2 eV and 1.5 eV incidence.}

%We further compare the THz response of \ce{Bi2Se3}(16\,QL)|CoFeB(4\,nm) with a Pt(16\,nm)|CoFeB(4\,nm) reference. Remarkably, the \textit{in-situ} grown TI samples outperform the Pt-based structure by approximately 50\%, underscoring the critical role of surface preservation in achieving strong THz emission. It is important to emphasize that this comparison reflects only the emitted THz signal, rather than the intrinsic SCC efficiency, since the high electrical conductivity of 16\,nm Pt significantly reduces THz out-coupling to free space. 

Next, we present the THz emission results obtained from \textit{in-situ} grown \ce{(Bi_{1-x}Sb_x)_2Te_3}(16\,QL)|CoFeB(4\,nm) samples. In the absence of the FM layer, the chemical potential in \ce{(Bi_{1-x}Sb_x)_2Te_3} can be tuned via Sb doping, approaching the Dirac point near $x \approx 0.9$ \cite{he2015scirep, kong2011natnano, zhang_band_2011, he_highly_2012, niu_realization_2012}. Figure~\ref{Fig3}\textbf{a} shows the time-domain THz emission from samples with $x = 0$, 0.5, 0.75, 0.9, and 1 under 1.55\,eV pump photon energy. We find that the THz emission amplitude reaches a minimum at $x \approx 0.9$, consistent with the charge neutrality point of the bare TI layer. This composition dependence is summarized in Fig.~\ref{Fig3}\textbf{b}. Notably, the stoichiometric \ce{Bi2Te3} sample exhibits THz emission efficiency comparable to that of \ce{Bi2Se3} (see Fig.~\ref{Fig1}). These results suggest that the SCC efficiency is strongly composition dependent. However, the precise correlation with the chemical potential remains unclear, as the presence of the FM layer may shift the Fermi level due to interfacial charge transfer \cite{zhang2016prb, culcer2010prb, eremeev2013prb, luo2013prb, hsu2017prb, spataru2014prb}. Whether the observed minimum in THz emission at $x \approx 0.9$ coincides with the intrinsic charge neutrality point of the TI layer or is merely coincidental requires further investigation. Repeating the measurement under a lower pump photon energy of 0.2\,eV yields the same trend (data not shown), suggesting the robustness of the observed composition dependence. In Fig.~\ref{Fig3}\textbf{c}, we compare \ce{Bi2Se3}(16\,QL)|CoFeB(4\,nm) and Pb-doped \ce{Bi2Se3}(16\,QL)|CoFeB(4\,nm). Without the FM layer, as-grown \ce{Bi2Se3} is bulk-conducting, while Pb doping makes it bulk-insulating. Similar to the (Bi,Sb)$_2$Te$_3$ case, we observe a significant reduction in THz emission from the Pb-doped sample, indicating suppressed SCC efficiency. Additional mechanisms, such as the surface warping effect \cite{Fu2009PRL, kong2011natnano, chen_experimental_2009}, and interfacial Rashba spin splitting \cite{Bianchi10, pauly2012prb, zhu2011prl, wang2015nanoletters} should also be considered for further investigation.

So far we have focused on the TI|FM structure. We finally explore the potential of SCC in WSM \cite{yan2017ar, wan2011prb, armitageRMP2018} using NbP. NbP is a type-I WSM, known to exhibit Weyl nodes and associated Fermi arcs in both bulk crystals and thin films, as confirmed by experimental studies \cite{souma2016prb, liu2016evolution, bedoyapinto2020acsnano, bedoya2021advmat}. We measure a bilayer composed of 5 nm NbP and 6 nm permalloy (Py). The time-domain THz signals under opposite magnetic fields are displayed in Fig.~\ref{Fig4}\textbf{a}. We find that the THz emission efficiency of this WSM|FM bilayer is about one-fifth of that in \textit{in-situ} grown \ce{Bi2Se3}(16 QL)|CoFeB(4 nm). In Fig.~\ref{Fig4}\textbf{b}, we present the THz electric field spectra obtained by performing a Fourier transform (FT) on the time-domain signals, taking into account the response function of a 250 $\mu$m-thick GaP EO crystal \cite{kampfrath2007sampling, leitenstorfer1999detectors}. 
%As references, we employ the conventional spin Hall THz emitters \cite{seifert2016NatPho}, W(2 nm)|CoFeB(1.8  nm)|Pt(2 nm) and CoFeB(4 nm)|Pt(3 nm), and 
We compare the spectra of the TI|FM samples \ce{Bi2Se3}(16 QL)|CoFeB(4 nm), \ce{(Bi_{0.1}Sb_{0.9})_2Te_3}(16 QL)|CoFeB(4 nm), and the WSM|FM sample NbP(5 nm)|Py(6 nm). We find that the majority of the spectral weight lies between 0.2 THz and 8.3 THz, with a cutoff around 8 THz set by GaP absorption. These broadband THz emission spectra of topological materials including TI and WSM are comparable to that from established spin Hall emitters. This underscores the potential of topological materials for efficient spintronic THz generation. We further emphasize the critical role of the metallic FM layer. It serves not only as the source of the spin-polarized current which boosts the THz emission, but also modifies the surface of the topological material, which may suppress SCC efficiency. Further studies are required to elucidate the interfacial mechanisms and optimize heterostructure design for maximized THz performance.\\

\begin{figure}
\centering
\includegraphics[width=0.5\textwidth]{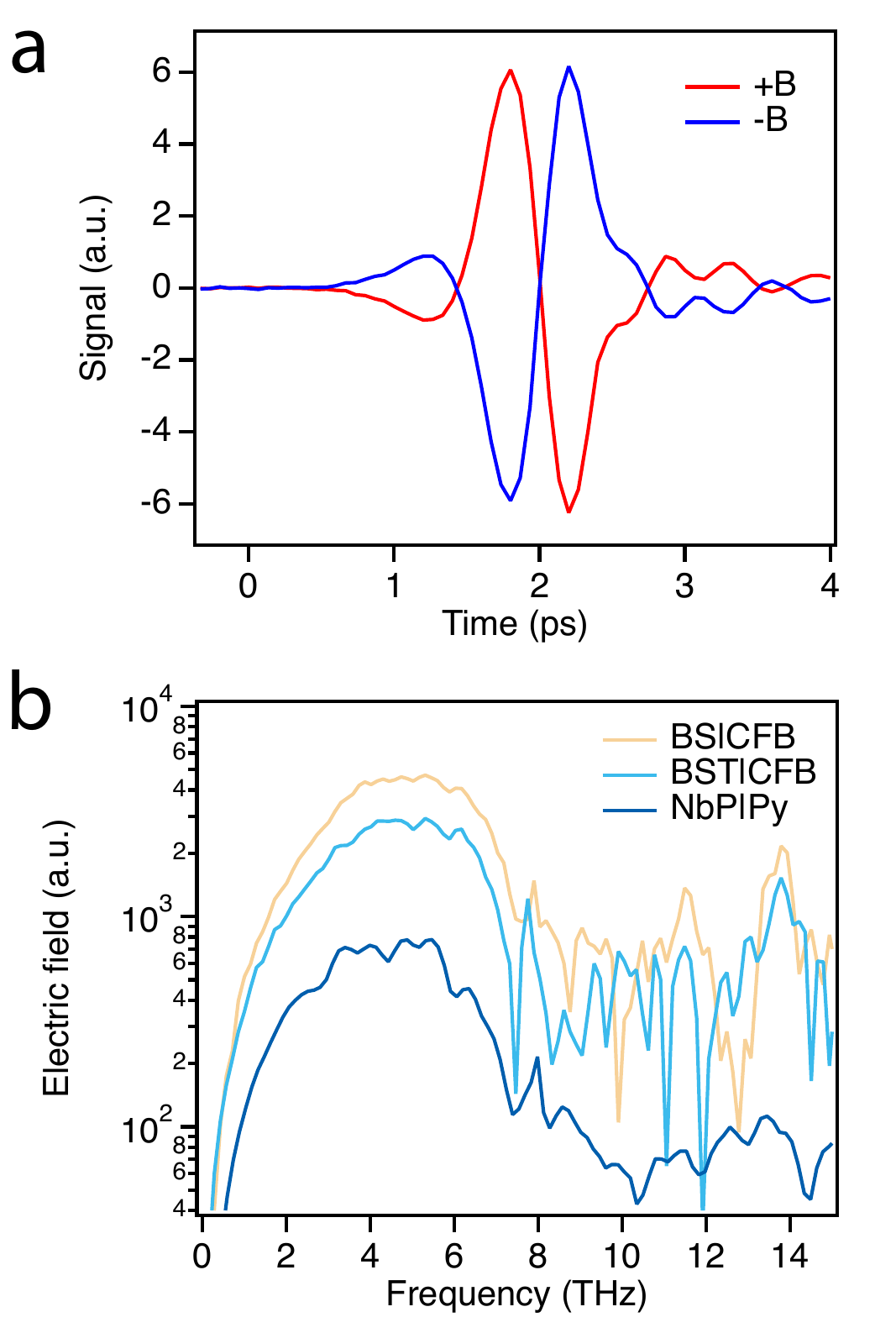}
\caption{
\textbf{a.} Time-domain THz waveforms for NbP(5 nm)|Py(6 nm) sample under opposite magnetic field directions. Pump photon energy: 0.8 eV.
\textbf{b.} Broadband THz emission spectra of different heterostructures. Pump photon energy: 1.55 eV.}
\label{Fig4}
\end{figure}

\textbf{Summary}

We have conducted a comprehensive study of THz emission in topological material|ferromagnetic metal (TM|FM) bilayer thin films. Our findings highlight the critical role of the TI layer condition without an exposure to the atmosphere in achieving high THz emission efficiency. In particular, preserving the TI surface from oxidation and optimizing its composition both significantly enhance the THz generation from spin to charger conversion. We further observe that the dependence on TI thickness varies with interface quality, indicating that thickness alone is not a reliable criterion for distinguishing between ISHE and IREE mechanisms. It is likely that both processes contribute to SCC in these heterostructures. Moreover, we demonstrate that WSM|FM bilayers can also facilitate efficient SCC, though the underlying mechanism requires further investigation. Finally, our broadband spectral measurements reveal that both TIs and WSMs can generate THz pulses with spectral bandwidths over 8 THz, underscoring the promise of topological materials for future ultrafast spintronic applications.

\textbf{Data and materials availability:} All data needed to evaluate the conclusions are present in the paper. Additional data related to this paper may be requested from the authors.\\

\textbf{Acknowledgments}
X.H. and L.W. acknowledge the support by Air Force Office of Scientific Research under award no. FA955022-1-0410. X.Y. and S.O. were sponsored by the Army Research Office and was accomplished under Grant Number W911NF-20-2-0166. X.H. acknowledges the Gordon and Betty Moore Foundation EPiQS Initiative through QuantEmX Travel Grant No. GBMF5305 to perform experiments in Freie Universität Berlin. The development of the THz emission spectroscopy was partially sponsored by the Army Research Office and was accomplished under Grant Number W911NF-25-2-0016. GB and TK acknowledge funding by the DFG Collaborative Research Center SFB TRR 227 “Ultrafast spin dynamics” (project ID 328545488, projects A05 and B02) and the ERC-2023 Advanced Grant ORBITERA (grant no. 101142285). A.B.-P. acknowledges support by the Generalitat Valenciana (CIDEGENT/2021/005).
\\

\textbf{Author Contributions:} L.W. conceived the project. X.H., Z.N. and L.W. built the THz emission setup. X.H. performed most of the THz emission experiments and analyzed the data with L.W. G.B. and C.I. performed the broadband THz emission experiments  under the supervision of T.K.. X.Y. and S.O. grew the TI films. T.R.T. and S.H. grew the FM films. A.B.-P. and S.S.P.P. grew the NbP bilayer film.\\

\textbf{Competing interests:} The authors declare that they have no competing interests.

\bibliography{Bi2Se3}

%merlin.mbs apsrev4-1.bst 2010-07-25 4.21a (PWD, AO, DPC) hacked
%Control: key (0)
%Control: author (8) initials jnrlst
%Control: editor formatted (1) identically to author
%Control: production of article title (-1) disabled
%Control: page (0) single
%Control: year (1) truncated
%Control: production of eprint (0) enabled
\begin{thebibliography}{64}%
\makeatletter
\providecommand \@ifxundefined [1]{%
 \@ifx{#1\undefined}
}%
\providecommand \@ifnum [1]{%
 \ifnum #1\expandafter \@firstoftwo
 \else \expandafter \@secondoftwo
 \fi
}%
\providecommand \@ifx [1]{%
 \ifx #1\expandafter \@firstoftwo
 \else \expandafter \@secondoftwo
 \fi
}%
\providecommand \natexlab [1]{#1}%
\providecommand \enquote  [1]{``#1''}%
\providecommand \bibnamefont  [1]{#1}%
\providecommand \bibfnamefont [1]{#1}%
\providecommand \citenamefont [1]{#1}%
\providecommand \href@noop [0]{\@secondoftwo}%
\providecommand \href [0]{\begingroup \@sanitize@url \@href}%
\providecommand \@href[1]{\@@startlink{#1}\@@href}%
\providecommand \@@href[1]{\endgroup#1\@@endlink}%
\providecommand \@sanitize@url [0]{\catcode `\\12\catcode `\$12\catcode
  `\&12\catcode `\#12\catcode `\^12\catcode `\_12\catcode `\%12\relax}%
\providecommand \@@startlink[1]{}%
\providecommand \@@endlink[0]{}%
\providecommand \url  [0]{\begingroup\@sanitize@url \@url }%
\providecommand \@url [1]{\endgroup\@href {#1}{\urlprefix }}%
\providecommand \urlprefix  [0]{URL }%
\providecommand \Eprint [0]{\href }%
\providecommand \doibase [0]{http://dx.doi.org/}%
\providecommand \selectlanguage [0]{\@gobble}%
\providecommand \bibinfo  [0]{\@secondoftwo}%
\providecommand \bibfield  [0]{\@secondoftwo}%
\providecommand \translation [1]{[#1]}%
\providecommand \BibitemOpen [0]{}%
\providecommand \bibitemStop [0]{}%
\providecommand \bibitemNoStop [0]{.\EOS\space}%
\providecommand \EOS [0]{\spacefactor3000\relax}%
\providecommand \BibitemShut  [1]{\csname bibitem#1\endcsname}%
\let\auto@bib@innerbib\@empty
%</preamble>
\bibitem [{\citenamefont {{\v{Z}}uti{\'c}}\ \emph {et~al.}(2004)\citenamefont
  {{\v{Z}}uti{\'c}}, \citenamefont {Fabian},\ and\ \citenamefont
  {Sarma}}]{zutic2004RMP}%
  \BibitemOpen
  \bibfield  {author} {\bibinfo {author} {\bibfnamefont {I.}~\bibnamefont
  {{\v{Z}}uti{\'c}}}, \bibinfo {author} {\bibfnamefont {J.}~\bibnamefont
  {Fabian}}, \ and\ \bibinfo {author} {\bibfnamefont {S.~D.}\ \bibnamefont
  {Sarma}},\ }\href@noop {} {\bibfield  {journal} {\bibinfo  {journal} {Reviews
  of modern physics}\ }\textbf {\bibinfo {volume} {76}},\ \bibinfo {pages}
  {323} (\bibinfo {year} {2004})}\BibitemShut {NoStop}%
\bibitem [{\citenamefont {Dieny}\ \emph {et~al.}(2020)\citenamefont {Dieny},
  \citenamefont {Prejbeanu}, \citenamefont {Garello}, \citenamefont
  {Gambardella}, \citenamefont {Freitas}, \citenamefont {Lehndorff},
  \citenamefont {Raberg}, \citenamefont {Ebels}, \citenamefont {Demokritov},
  \citenamefont {Akerman} \emph {et~al.}}]{dieny2020natelec}%
  \BibitemOpen
  \bibfield  {author} {\bibinfo {author} {\bibfnamefont {B.}~\bibnamefont
  {Dieny}}, \bibinfo {author} {\bibfnamefont {I.~L.}\ \bibnamefont
  {Prejbeanu}}, \bibinfo {author} {\bibfnamefont {K.}~\bibnamefont {Garello}},
  \bibinfo {author} {\bibfnamefont {P.}~\bibnamefont {Gambardella}}, \bibinfo
  {author} {\bibfnamefont {P.}~\bibnamefont {Freitas}}, \bibinfo {author}
  {\bibfnamefont {R.}~\bibnamefont {Lehndorff}}, \bibinfo {author}
  {\bibfnamefont {W.}~\bibnamefont {Raberg}}, \bibinfo {author} {\bibfnamefont
  {U.}~\bibnamefont {Ebels}}, \bibinfo {author} {\bibfnamefont {S.~O.}\
  \bibnamefont {Demokritov}}, \bibinfo {author} {\bibfnamefont
  {J.}~\bibnamefont {Akerman}},  \emph {et~al.},\ }\href@noop {} {\bibfield
  {journal} {\bibinfo  {journal} {Nature Electronics}\ }\textbf {\bibinfo
  {volume} {3}},\ \bibinfo {pages} {446} (\bibinfo {year} {2020})}\BibitemShut
  {NoStop}%
\bibitem [{\citenamefont {Han}\ \emph {et~al.}(2018)\citenamefont {Han},
  \citenamefont {Otani},\ and\ \citenamefont {Maekawa}}]{han2018npjquan}%
  \BibitemOpen
  \bibfield  {author} {\bibinfo {author} {\bibfnamefont {W.}~\bibnamefont
  {Han}}, \bibinfo {author} {\bibfnamefont {Y.}~\bibnamefont {Otani}}, \ and\
  \bibinfo {author} {\bibfnamefont {S.}~\bibnamefont {Maekawa}},\ }\href@noop
  {} {\bibfield  {journal} {\bibinfo  {journal} {npj Quantum Materials}\
  }\textbf {\bibinfo {volume} {3}},\ \bibinfo {pages} {27} (\bibinfo {year}
  {2018})}\BibitemShut {NoStop}%
\bibitem [{\citenamefont {S{\'a}nchez}\ \emph {et~al.}(2013)\citenamefont
  {S{\'a}nchez}, \citenamefont {Vila}, \citenamefont {Desfonds}, \citenamefont
  {Gambarelli}, \citenamefont {Attan{\'e}}, \citenamefont {De~Teresa},
  \citenamefont {Mag{\'e}n},\ and\ \citenamefont {Fert}}]{sanchez2013NatCom}%
  \BibitemOpen
  \bibfield  {author} {\bibinfo {author} {\bibfnamefont {J.~R.}\ \bibnamefont
  {S{\'a}nchez}}, \bibinfo {author} {\bibfnamefont {L.}~\bibnamefont {Vila}},
  \bibinfo {author} {\bibfnamefont {G.}~\bibnamefont {Desfonds}}, \bibinfo
  {author} {\bibfnamefont {S.}~\bibnamefont {Gambarelli}}, \bibinfo {author}
  {\bibfnamefont {J.}~\bibnamefont {Attan{\'e}}}, \bibinfo {author}
  {\bibfnamefont {J.}~\bibnamefont {De~Teresa}}, \bibinfo {author}
  {\bibfnamefont {C.}~\bibnamefont {Mag{\'e}n}}, \ and\ \bibinfo {author}
  {\bibfnamefont {A.}~\bibnamefont {Fert}},\ }\href@noop {} {\bibfield
  {journal} {\bibinfo  {journal} {Nature communications}\ }\textbf {\bibinfo
  {volume} {4}},\ \bibinfo {pages} {2944} (\bibinfo {year} {2013})}\BibitemShut
  {NoStop}%
\bibitem [{\citenamefont {Sinova}\ \emph {et~al.}(2015)\citenamefont {Sinova},
  \citenamefont {Valenzuela}, \citenamefont {Wunderlich}, \citenamefont
  {Back},\ and\ \citenamefont {Jungwirth}}]{sinova2015RMP}%
  \BibitemOpen
  \bibfield  {author} {\bibinfo {author} {\bibfnamefont {J.}~\bibnamefont
  {Sinova}}, \bibinfo {author} {\bibfnamefont {S.~O.}\ \bibnamefont
  {Valenzuela}}, \bibinfo {author} {\bibfnamefont {J.}~\bibnamefont
  {Wunderlich}}, \bibinfo {author} {\bibfnamefont {C.}~\bibnamefont {Back}}, \
  and\ \bibinfo {author} {\bibfnamefont {T.}~\bibnamefont {Jungwirth}},\
  }\href@noop {} {\bibfield  {journal} {\bibinfo  {journal} {Reviews of modern
  physics}\ }\textbf {\bibinfo {volume} {87}},\ \bibinfo {pages} {1213}
  (\bibinfo {year} {2015})}\BibitemShut {NoStop}%
\bibitem [{\citenamefont {Edelstein}(1990)}]{EDELSTEIN1990ssc}%
  \BibitemOpen
  \bibfield  {author} {\bibinfo {author} {\bibfnamefont {V.~M.}\ \bibnamefont
  {Edelstein}},\ }\href@noop {} {\bibfield  {journal} {\bibinfo  {journal}
  {Solid State Communications}\ }\textbf {\bibinfo {volume} {73}},\ \bibinfo
  {pages} {233} (\bibinfo {year} {1990})}\BibitemShut {NoStop}%
\bibitem [{\citenamefont {Seifert}\ \emph {et~al.}(2016)\citenamefont
  {Seifert}, \citenamefont {Jaiswal}, \citenamefont {Martens}, \citenamefont
  {Hannegan}, \citenamefont {Braun}, \citenamefont {Maldonado}, \citenamefont
  {Freimuth}, \citenamefont {Kronenberg}, \citenamefont {Henrizi},
  \citenamefont {Radu} \emph {et~al.}}]{seifert2016NatPho}%
  \BibitemOpen
  \bibfield  {author} {\bibinfo {author} {\bibfnamefont {T.}~\bibnamefont
  {Seifert}}, \bibinfo {author} {\bibfnamefont {S.}~\bibnamefont {Jaiswal}},
  \bibinfo {author} {\bibfnamefont {U.}~\bibnamefont {Martens}}, \bibinfo
  {author} {\bibfnamefont {J.}~\bibnamefont {Hannegan}}, \bibinfo {author}
  {\bibfnamefont {L.}~\bibnamefont {Braun}}, \bibinfo {author} {\bibfnamefont
  {P.}~\bibnamefont {Maldonado}}, \bibinfo {author} {\bibfnamefont
  {F.}~\bibnamefont {Freimuth}}, \bibinfo {author} {\bibfnamefont
  {A.}~\bibnamefont {Kronenberg}}, \bibinfo {author} {\bibfnamefont
  {J.}~\bibnamefont {Henrizi}}, \bibinfo {author} {\bibfnamefont
  {I.}~\bibnamefont {Radu}},  \emph {et~al.},\ }\href@noop {} {\bibfield
  {journal} {\bibinfo  {journal} {Nature photonics}\ }\textbf {\bibinfo
  {volume} {10}},\ \bibinfo {pages} {483} (\bibinfo {year} {2016})}\BibitemShut
  {NoStop}%
\bibitem [{\citenamefont {Zhu}\ \emph {et~al.}(2015)\citenamefont {Zhu},
  \citenamefont {Kubera}, \citenamefont {Fai~Mak},\ and\ \citenamefont
  {Shan}}]{zhu2015SciRep}%
  \BibitemOpen
  \bibfield  {author} {\bibinfo {author} {\bibfnamefont {L.-G.}\ \bibnamefont
  {Zhu}}, \bibinfo {author} {\bibfnamefont {B.}~\bibnamefont {Kubera}},
  \bibinfo {author} {\bibfnamefont {K.}~\bibnamefont {Fai~Mak}}, \ and\
  \bibinfo {author} {\bibfnamefont {J.}~\bibnamefont {Shan}},\ }\href@noop {}
  {\bibfield  {journal} {\bibinfo  {journal} {Scientific reports}\ }\textbf
  {\bibinfo {volume} {5}},\ \bibinfo {pages} {10308} (\bibinfo {year}
  {2015})}\BibitemShut {NoStop}%
\bibitem [{\citenamefont {Wang}\ \emph {et~al.}(2018)\citenamefont {Wang},
  \citenamefont {Cheng}, \citenamefont {Zhu}, \citenamefont {Wu}, \citenamefont
  {Chen}, \citenamefont {Wang}, \citenamefont {Zhao}, \citenamefont
  {Boothroyd}, \citenamefont {Lam}, \citenamefont {Zhu} \emph
  {et~al.}}]{wang2018AdvMat}%
  \BibitemOpen
  \bibfield  {author} {\bibinfo {author} {\bibfnamefont {X.}~\bibnamefont
  {Wang}}, \bibinfo {author} {\bibfnamefont {L.}~\bibnamefont {Cheng}},
  \bibinfo {author} {\bibfnamefont {D.}~\bibnamefont {Zhu}}, \bibinfo {author}
  {\bibfnamefont {Y.}~\bibnamefont {Wu}}, \bibinfo {author} {\bibfnamefont
  {M.}~\bibnamefont {Chen}}, \bibinfo {author} {\bibfnamefont {Y.}~\bibnamefont
  {Wang}}, \bibinfo {author} {\bibfnamefont {D.}~\bibnamefont {Zhao}}, \bibinfo
  {author} {\bibfnamefont {C.~B.}\ \bibnamefont {Boothroyd}}, \bibinfo {author}
  {\bibfnamefont {Y.~M.}\ \bibnamefont {Lam}}, \bibinfo {author} {\bibfnamefont
  {J.-X.}\ \bibnamefont {Zhu}},  \emph {et~al.},\ }\href@noop {} {\bibfield
  {journal} {\bibinfo  {journal} {Advanced Materials}\ }\textbf {\bibinfo
  {volume} {30}},\ \bibinfo {pages} {1802356} (\bibinfo {year}
  {2018})}\BibitemShut {NoStop}%
\bibitem [{\citenamefont {Wang}\ \emph {et~al.}(2017)\citenamefont {Wang},
  \citenamefont {Zhu}, \citenamefont {Wu}, \citenamefont {Yang}, \citenamefont
  {Yu}, \citenamefont {Ramaswamy}, \citenamefont {Mishra}, \citenamefont {Shi},
  \citenamefont {Elyasi}, \citenamefont {Teo} \emph
  {et~al.}}]{wang2017natcomm}%
  \BibitemOpen
  \bibfield  {author} {\bibinfo {author} {\bibfnamefont {Y.}~\bibnamefont
  {Wang}}, \bibinfo {author} {\bibfnamefont {D.}~\bibnamefont {Zhu}}, \bibinfo
  {author} {\bibfnamefont {Y.}~\bibnamefont {Wu}}, \bibinfo {author}
  {\bibfnamefont {Y.}~\bibnamefont {Yang}}, \bibinfo {author} {\bibfnamefont
  {J.}~\bibnamefont {Yu}}, \bibinfo {author} {\bibfnamefont {R.}~\bibnamefont
  {Ramaswamy}}, \bibinfo {author} {\bibfnamefont {R.}~\bibnamefont {Mishra}},
  \bibinfo {author} {\bibfnamefont {S.}~\bibnamefont {Shi}}, \bibinfo {author}
  {\bibfnamefont {M.}~\bibnamefont {Elyasi}}, \bibinfo {author} {\bibfnamefont
  {K.-L.}\ \bibnamefont {Teo}},  \emph {et~al.},\ }\href@noop {} {\bibfield
  {journal} {\bibinfo  {journal} {Nature communications}\ }\textbf {\bibinfo
  {volume} {8}},\ \bibinfo {pages} {1364} (\bibinfo {year} {2017})}\BibitemShut
  {NoStop}%
\bibitem [{\citenamefont {Noyan}\ \emph {et~al.}(2025)\citenamefont {Noyan},
  \citenamefont {Zhang}, \citenamefont {Moon}, \citenamefont {Cobas},
  \citenamefont {Lohmann}, \citenamefont {Zou}, \citenamefont {Li},
  \citenamefont {Weinert}, \citenamefont {Jonker}, \citenamefont {van~‘t
  Erve} \emph {et~al.}}]{noyan2025nanoletters}%
  \BibitemOpen
  \bibfield  {author} {\bibinfo {author} {\bibfnamefont {M.~A.}\ \bibnamefont
  {Noyan}}, \bibinfo {author} {\bibfnamefont {X.}~\bibnamefont {Zhang}},
  \bibinfo {author} {\bibfnamefont {J.}~\bibnamefont {Moon}}, \bibinfo {author}
  {\bibfnamefont {E.}~\bibnamefont {Cobas}}, \bibinfo {author} {\bibfnamefont
  {M.}~\bibnamefont {Lohmann}}, \bibinfo {author} {\bibfnamefont
  {Q.}~\bibnamefont {Zou}}, \bibinfo {author} {\bibfnamefont {L.}~\bibnamefont
  {Li}}, \bibinfo {author} {\bibfnamefont {M.}~\bibnamefont {Weinert}},
  \bibinfo {author} {\bibfnamefont {B.~T.}\ \bibnamefont {Jonker}}, \bibinfo
  {author} {\bibfnamefont {O.~M.}\ \bibnamefont {van~‘t Erve}},  \emph
  {et~al.},\ }\href@noop {} {\bibfield  {journal} {\bibinfo  {journal} {Nano
  Letters}\ } (\bibinfo {year} {2025})}\BibitemShut {NoStop}%
\bibitem [{\citenamefont {Shiomi}\ \emph {et~al.}(2014)\citenamefont {Shiomi},
  \citenamefont {Nomura}, \citenamefont {Kajiwara}, \citenamefont {Eto},
  \citenamefont {Novak}, \citenamefont {Segawa}, \citenamefont {Ando},\ and\
  \citenamefont {Saitoh}}]{shiomi2014prl}%
  \BibitemOpen
  \bibfield  {author} {\bibinfo {author} {\bibfnamefont {Y.}~\bibnamefont
  {Shiomi}}, \bibinfo {author} {\bibfnamefont {K.}~\bibnamefont {Nomura}},
  \bibinfo {author} {\bibfnamefont {Y.}~\bibnamefont {Kajiwara}}, \bibinfo
  {author} {\bibfnamefont {K.}~\bibnamefont {Eto}}, \bibinfo {author}
  {\bibfnamefont {M.}~\bibnamefont {Novak}}, \bibinfo {author} {\bibfnamefont
  {K.}~\bibnamefont {Segawa}}, \bibinfo {author} {\bibfnamefont
  {Y.}~\bibnamefont {Ando}}, \ and\ \bibinfo {author} {\bibfnamefont
  {E.}~\bibnamefont {Saitoh}},\ }\href {\doibase
  10.1103/PhysRevLett.113.196601} {\bibfield  {journal} {\bibinfo  {journal}
  {Phys. Rev. Lett.}\ }\textbf {\bibinfo {volume} {113}},\ \bibinfo {pages}
  {196601} (\bibinfo {year} {2014})}\BibitemShut {NoStop}%
\bibitem [{\citenamefont {Mellnik}\ \emph {et~al.}(2014)\citenamefont
  {Mellnik}, \citenamefont {Lee}, \citenamefont {Richardella}, \citenamefont
  {Grab}, \citenamefont {Mintun}, \citenamefont {Fischer}, \citenamefont
  {Vaezi}, \citenamefont {Manchon}, \citenamefont {Kim}, \citenamefont
  {Samarth} \emph {et~al.}}]{mellnik2014nature}%
  \BibitemOpen
  \bibfield  {author} {\bibinfo {author} {\bibfnamefont {A.~R.}\ \bibnamefont
  {Mellnik}}, \bibinfo {author} {\bibfnamefont {J.}~\bibnamefont {Lee}},
  \bibinfo {author} {\bibfnamefont {A.}~\bibnamefont {Richardella}}, \bibinfo
  {author} {\bibfnamefont {J.~L.}\ \bibnamefont {Grab}}, \bibinfo {author}
  {\bibfnamefont {P.~J.}\ \bibnamefont {Mintun}}, \bibinfo {author}
  {\bibfnamefont {M.~H.}\ \bibnamefont {Fischer}}, \bibinfo {author}
  {\bibfnamefont {A.}~\bibnamefont {Vaezi}}, \bibinfo {author} {\bibfnamefont
  {A.}~\bibnamefont {Manchon}}, \bibinfo {author} {\bibfnamefont {E.-A.}\
  \bibnamefont {Kim}}, \bibinfo {author} {\bibfnamefont {N.}~\bibnamefont
  {Samarth}},  \emph {et~al.},\ }\href@noop {} {\bibfield  {journal} {\bibinfo
  {journal} {Nature}\ }\textbf {\bibinfo {volume} {511}},\ \bibinfo {pages}
  {449} (\bibinfo {year} {2014})}\BibitemShut {NoStop}%
\bibitem [{\citenamefont {Wang}\ \emph {et~al.}(2016)\citenamefont {Wang},
  \citenamefont {Kally}, \citenamefont {Lee}, \citenamefont {Liu},
  \citenamefont {Chang}, \citenamefont {Hickey}, \citenamefont {Mkhoyan},
  \citenamefont {Wu}, \citenamefont {Richardella},\ and\ \citenamefont
  {Samarth}}]{wang2016prl}%
  \BibitemOpen
  \bibfield  {author} {\bibinfo {author} {\bibfnamefont {H.}~\bibnamefont
  {Wang}}, \bibinfo {author} {\bibfnamefont {J.}~\bibnamefont {Kally}},
  \bibinfo {author} {\bibfnamefont {J.~S.}\ \bibnamefont {Lee}}, \bibinfo
  {author} {\bibfnamefont {T.}~\bibnamefont {Liu}}, \bibinfo {author}
  {\bibfnamefont {H.}~\bibnamefont {Chang}}, \bibinfo {author} {\bibfnamefont
  {D.~R.}\ \bibnamefont {Hickey}}, \bibinfo {author} {\bibfnamefont {K.~A.}\
  \bibnamefont {Mkhoyan}}, \bibinfo {author} {\bibfnamefont {M.}~\bibnamefont
  {Wu}}, \bibinfo {author} {\bibfnamefont {A.}~\bibnamefont {Richardella}}, \
  and\ \bibinfo {author} {\bibfnamefont {N.}~\bibnamefont {Samarth}},\ }\href
  {\doibase 10.1103/PhysRevLett.117.076601} {\bibfield  {journal} {\bibinfo
  {journal} {Phys. Rev. Lett.}\ }\textbf {\bibinfo {volume} {117}},\ \bibinfo
  {pages} {076601} (\bibinfo {year} {2016})}\BibitemShut {NoStop}%
\bibitem [{\citenamefont {Lee}\ \emph {et~al.}(2015)\citenamefont {Lee},
  \citenamefont {Richardella}, \citenamefont {Hickey}, \citenamefont
  {Mkhoyan},\ and\ \citenamefont {Samarth}}]{lee2015prb}%
  \BibitemOpen
  \bibfield  {author} {\bibinfo {author} {\bibfnamefont {J.~S.}\ \bibnamefont
  {Lee}}, \bibinfo {author} {\bibfnamefont {A.}~\bibnamefont {Richardella}},
  \bibinfo {author} {\bibfnamefont {D.~R.}\ \bibnamefont {Hickey}}, \bibinfo
  {author} {\bibfnamefont {K.~A.}\ \bibnamefont {Mkhoyan}}, \ and\ \bibinfo
  {author} {\bibfnamefont {N.}~\bibnamefont {Samarth}},\ }\href@noop {}
  {\bibfield  {journal} {\bibinfo  {journal} {Physical Review B}\ }\textbf
  {\bibinfo {volume} {92}},\ \bibinfo {pages} {155312} (\bibinfo {year}
  {2015})}\BibitemShut {NoStop}%
\bibitem [{\citenamefont {Deorani}\ \emph {et~al.}(2014)\citenamefont
  {Deorani}, \citenamefont {Son}, \citenamefont {Banerjee}, \citenamefont
  {Koirala}, \citenamefont {Brahlek}, \citenamefont {Oh},\ and\ \citenamefont
  {Yang}}]{deorani2014prb}%
  \BibitemOpen
  \bibfield  {author} {\bibinfo {author} {\bibfnamefont {P.}~\bibnamefont
  {Deorani}}, \bibinfo {author} {\bibfnamefont {J.}~\bibnamefont {Son}},
  \bibinfo {author} {\bibfnamefont {K.}~\bibnamefont {Banerjee}}, \bibinfo
  {author} {\bibfnamefont {N.}~\bibnamefont {Koirala}}, \bibinfo {author}
  {\bibfnamefont {M.}~\bibnamefont {Brahlek}}, \bibinfo {author} {\bibfnamefont
  {S.}~\bibnamefont {Oh}}, \ and\ \bibinfo {author} {\bibfnamefont
  {H.}~\bibnamefont {Yang}},\ }\href@noop {} {\bibfield  {journal} {\bibinfo
  {journal} {Physical Review B}\ }\textbf {\bibinfo {volume} {90}},\ \bibinfo
  {pages} {094403} (\bibinfo {year} {2014})}\BibitemShut {NoStop}%
\bibitem [{\citenamefont {Jamali}\ \emph {et~al.}(2015)\citenamefont {Jamali},
  \citenamefont {Lee}, \citenamefont {Jeong}, \citenamefont {Mahfouzi},
  \citenamefont {Lv}, \citenamefont {Zhao}, \citenamefont {Nikolic},
  \citenamefont {Mkhoyan}, \citenamefont {Samarth},\ and\ \citenamefont
  {Wang}}]{jamali2015nl}%
  \BibitemOpen
  \bibfield  {author} {\bibinfo {author} {\bibfnamefont {M.}~\bibnamefont
  {Jamali}}, \bibinfo {author} {\bibfnamefont {J.~S.}\ \bibnamefont {Lee}},
  \bibinfo {author} {\bibfnamefont {J.~S.}\ \bibnamefont {Jeong}}, \bibinfo
  {author} {\bibfnamefont {F.}~\bibnamefont {Mahfouzi}}, \bibinfo {author}
  {\bibfnamefont {Y.}~\bibnamefont {Lv}}, \bibinfo {author} {\bibfnamefont
  {Z.}~\bibnamefont {Zhao}}, \bibinfo {author} {\bibfnamefont {B.~K.}\
  \bibnamefont {Nikolic}}, \bibinfo {author} {\bibfnamefont {K.~A.}\
  \bibnamefont {Mkhoyan}}, \bibinfo {author} {\bibfnamefont {N.}~\bibnamefont
  {Samarth}}, \ and\ \bibinfo {author} {\bibfnamefont {J.-P.}\ \bibnamefont
  {Wang}},\ }\href@noop {} {\bibfield  {journal} {\bibinfo  {journal} {Nano
  letters}\ }\textbf {\bibinfo {volume} {15}},\ \bibinfo {pages} {7126}
  (\bibinfo {year} {2015})}\BibitemShut {NoStop}%
\bibitem [{\citenamefont {Wang}\ \emph {et~al.}(2019)\citenamefont {Wang},
  \citenamefont {Kally}, \citenamefont {Sahin}, \citenamefont {Liu},
  \citenamefont {Yanez}, \citenamefont {Kamp}, \citenamefont {Richardella},
  \citenamefont {Wu}, \citenamefont {Flatt{\'e}},\ and\ \citenamefont
  {Samarth}}]{wang2019prr}%
  \BibitemOpen
  \bibfield  {author} {\bibinfo {author} {\bibfnamefont {H.}~\bibnamefont
  {Wang}}, \bibinfo {author} {\bibfnamefont {J.}~\bibnamefont {Kally}},
  \bibinfo {author} {\bibfnamefont {C.}~\bibnamefont {Sahin}}, \bibinfo
  {author} {\bibfnamefont {T.}~\bibnamefont {Liu}}, \bibinfo {author}
  {\bibfnamefont {W.}~\bibnamefont {Yanez}}, \bibinfo {author} {\bibfnamefont
  {E.~J.}\ \bibnamefont {Kamp}}, \bibinfo {author} {\bibfnamefont
  {A.}~\bibnamefont {Richardella}}, \bibinfo {author} {\bibfnamefont
  {M.}~\bibnamefont {Wu}}, \bibinfo {author} {\bibfnamefont {M.~E.}\
  \bibnamefont {Flatt{\'e}}}, \ and\ \bibinfo {author} {\bibfnamefont
  {N.}~\bibnamefont {Samarth}},\ }\href@noop {} {\bibfield  {journal} {\bibinfo
   {journal} {Physical review research}\ }\textbf {\bibinfo {volume} {1}},\
  \bibinfo {pages} {012014} (\bibinfo {year} {2019})}\BibitemShut {NoStop}%
\bibitem [{\citenamefont {Sahu}\ \emph {et~al.}(2023)\citenamefont {Sahu},
  \citenamefont {Yang}, \citenamefont {Fan}, \citenamefont {Jaffr{\`e}s},
  \citenamefont {Chen}, \citenamefont {Devaux}, \citenamefont {Fagot-Revurat},
  \citenamefont {Migot}, \citenamefont {Rongione}, \citenamefont {Chen} \emph
  {et~al.}}]{sahu2023acsami}%
  \BibitemOpen
  \bibfield  {author} {\bibinfo {author} {\bibfnamefont {P.}~\bibnamefont
  {Sahu}}, \bibinfo {author} {\bibfnamefont {Y.}~\bibnamefont {Yang}}, \bibinfo
  {author} {\bibfnamefont {Y.}~\bibnamefont {Fan}}, \bibinfo {author}
  {\bibfnamefont {H.}~\bibnamefont {Jaffr{\`e}s}}, \bibinfo {author}
  {\bibfnamefont {J.-Y.}\ \bibnamefont {Chen}}, \bibinfo {author}
  {\bibfnamefont {X.}~\bibnamefont {Devaux}}, \bibinfo {author} {\bibfnamefont
  {Y.}~\bibnamefont {Fagot-Revurat}}, \bibinfo {author} {\bibfnamefont
  {S.}~\bibnamefont {Migot}}, \bibinfo {author} {\bibfnamefont
  {E.}~\bibnamefont {Rongione}}, \bibinfo {author} {\bibfnamefont
  {T.}~\bibnamefont {Chen}},  \emph {et~al.},\ }\href@noop {} {\bibfield
  {journal} {\bibinfo  {journal} {ACS Applied Materials \& Interfaces}\
  }\textbf {\bibinfo {volume} {15}},\ \bibinfo {pages} {38592} (\bibinfo {year}
  {2023})}\BibitemShut {NoStop}%
\bibitem [{\citenamefont {Shi}\ \emph {et~al.}(2025)\citenamefont {Shi},
  \citenamefont {Liu}, \citenamefont {Hu}, \citenamefont {Shi}, \citenamefont
  {Manchon},\ and\ \citenamefont {Yang}}]{shi2025prb}%
  \BibitemOpen
  \bibfield  {author} {\bibinfo {author} {\bibfnamefont {S.}~\bibnamefont
  {Shi}}, \bibinfo {author} {\bibfnamefont {E.}~\bibnamefont {Liu}}, \bibinfo
  {author} {\bibfnamefont {F.}~\bibnamefont {Hu}}, \bibinfo {author}
  {\bibfnamefont {G.}~\bibnamefont {Shi}}, \bibinfo {author} {\bibfnamefont
  {A.}~\bibnamefont {Manchon}}, \ and\ \bibinfo {author} {\bibfnamefont
  {H.}~\bibnamefont {Yang}},\ }\href@noop {} {\bibfield  {journal} {\bibinfo
  {journal} {Physical Review B}\ }\textbf {\bibinfo {volume} {111}},\ \bibinfo
  {pages} {094433} (\bibinfo {year} {2025})}\BibitemShut {NoStop}%
\bibitem [{\citenamefont {Ni}\ \emph {et~al.}(2021)\citenamefont {Ni},
  \citenamefont {Wang}, \citenamefont {Zhang}, \citenamefont {Pozo},
  \citenamefont {Xu}, \citenamefont {Han}, \citenamefont {Manna}, \citenamefont
  {Paglione}, \citenamefont {Felser}, \citenamefont {Grushin} \emph
  {et~al.}}]{ni2021NatCom}%
  \BibitemOpen
  \bibfield  {author} {\bibinfo {author} {\bibfnamefont {Z.}~\bibnamefont
  {Ni}}, \bibinfo {author} {\bibfnamefont {K.}~\bibnamefont {Wang}}, \bibinfo
  {author} {\bibfnamefont {Y.}~\bibnamefont {Zhang}}, \bibinfo {author}
  {\bibfnamefont {O.}~\bibnamefont {Pozo}}, \bibinfo {author} {\bibfnamefont
  {B.}~\bibnamefont {Xu}}, \bibinfo {author} {\bibfnamefont {X.}~\bibnamefont
  {Han}}, \bibinfo {author} {\bibfnamefont {K.}~\bibnamefont {Manna}}, \bibinfo
  {author} {\bibfnamefont {J.}~\bibnamefont {Paglione}}, \bibinfo {author}
  {\bibfnamefont {C.}~\bibnamefont {Felser}}, \bibinfo {author} {\bibfnamefont
  {A.~G.}\ \bibnamefont {Grushin}},  \emph {et~al.},\ }\href@noop {} {\bibfield
   {journal} {\bibinfo  {journal} {Nature communications}\ }\textbf {\bibinfo
  {volume} {12}},\ \bibinfo {pages} {1} (\bibinfo {year} {2021})}\BibitemShut
  {NoStop}%
\bibitem [{\citenamefont {Rouzegar}\ \emph {et~al.}(2022)\citenamefont
  {Rouzegar}, \citenamefont {Brandt}, \citenamefont {N{\'a}dvorn{\'\i}k},
  \citenamefont {Reiss}, \citenamefont {Chekhov}, \citenamefont {Gueckstock},
  \citenamefont {In}, \citenamefont {Wolf}, \citenamefont {Seifert},
  \citenamefont {Brouwer} \emph {et~al.}}]{rouzegar2022prb}%
  \BibitemOpen
  \bibfield  {author} {\bibinfo {author} {\bibfnamefont {R.}~\bibnamefont
  {Rouzegar}}, \bibinfo {author} {\bibfnamefont {L.}~\bibnamefont {Brandt}},
  \bibinfo {author} {\bibfnamefont {L.}~\bibnamefont {N{\'a}dvorn{\'\i}k}},
  \bibinfo {author} {\bibfnamefont {D.~A.}\ \bibnamefont {Reiss}}, \bibinfo
  {author} {\bibfnamefont {A.~L.}\ \bibnamefont {Chekhov}}, \bibinfo {author}
  {\bibfnamefont {O.}~\bibnamefont {Gueckstock}}, \bibinfo {author}
  {\bibfnamefont {C.}~\bibnamefont {In}}, \bibinfo {author} {\bibfnamefont
  {M.}~\bibnamefont {Wolf}}, \bibinfo {author} {\bibfnamefont {T.~S.}\
  \bibnamefont {Seifert}}, \bibinfo {author} {\bibfnamefont {P.~W.}\
  \bibnamefont {Brouwer}},  \emph {et~al.},\ }\href@noop {} {\bibfield
  {journal} {\bibinfo  {journal} {Physical Review B}\ }\textbf {\bibinfo
  {volume} {106}},\ \bibinfo {pages} {144427} (\bibinfo {year}
  {2022})}\BibitemShut {NoStop}%
\bibitem [{\citenamefont {Beaurepaire}\ \emph {et~al.}(2004)\citenamefont
  {Beaurepaire}, \citenamefont {Turner}, \citenamefont {Harrel}, \citenamefont
  {Beard}, \citenamefont {Bigot},\ and\ \citenamefont
  {Schmuttenmaer}}]{beaurepaire2004apl}%
  \BibitemOpen
  \bibfield  {author} {\bibinfo {author} {\bibfnamefont {E.}~\bibnamefont
  {Beaurepaire}}, \bibinfo {author} {\bibfnamefont {G.}~\bibnamefont {Turner}},
  \bibinfo {author} {\bibfnamefont {S.}~\bibnamefont {Harrel}}, \bibinfo
  {author} {\bibfnamefont {M.}~\bibnamefont {Beard}}, \bibinfo {author}
  {\bibfnamefont {J.-Y.}\ \bibnamefont {Bigot}}, \ and\ \bibinfo {author}
  {\bibfnamefont {C.}~\bibnamefont {Schmuttenmaer}},\ }\href@noop {} {\bibfield
   {journal} {\bibinfo  {journal} {Applied Physics Letters}\ }\textbf {\bibinfo
  {volume} {84}},\ \bibinfo {pages} {3465} (\bibinfo {year}
  {2004})}\BibitemShut {NoStop}%
\bibitem [{\citenamefont {Zhang}\ \emph {et~al.}(2016)\citenamefont {Zhang},
  \citenamefont {Velev}, \citenamefont {Dang},\ and\ \citenamefont
  {Tsymbal}}]{zhang2016prb}%
  \BibitemOpen
  \bibfield  {author} {\bibinfo {author} {\bibfnamefont {J.}~\bibnamefont
  {Zhang}}, \bibinfo {author} {\bibfnamefont {J.~P.}\ \bibnamefont {Velev}},
  \bibinfo {author} {\bibfnamefont {X.}~\bibnamefont {Dang}}, \ and\ \bibinfo
  {author} {\bibfnamefont {E.~Y.}\ \bibnamefont {Tsymbal}},\ }\href@noop {}
  {\bibfield  {journal} {\bibinfo  {journal} {Physical Review B}\ }\textbf
  {\bibinfo {volume} {94}},\ \bibinfo {pages} {014435} (\bibinfo {year}
  {2016})}\BibitemShut {NoStop}%
\bibitem [{\citenamefont {Culcer}\ \emph {et~al.}(2010)\citenamefont {Culcer},
  \citenamefont {Hwang}, \citenamefont {Stanescu},\ and\ \citenamefont
  {Das~Sarma}}]{culcer2010prb}%
  \BibitemOpen
  \bibfield  {author} {\bibinfo {author} {\bibfnamefont {D.}~\bibnamefont
  {Culcer}}, \bibinfo {author} {\bibfnamefont {E.~H.}\ \bibnamefont {Hwang}},
  \bibinfo {author} {\bibfnamefont {T.~D.}\ \bibnamefont {Stanescu}}, \ and\
  \bibinfo {author} {\bibfnamefont {S.}~\bibnamefont {Das~Sarma}},\ }\href
  {\doibase 10.1103/PhysRevB.82.155457} {\bibfield  {journal} {\bibinfo
  {journal} {Phys. Rev. B}\ }\textbf {\bibinfo {volume} {82}},\ \bibinfo
  {pages} {155457} (\bibinfo {year} {2010})}\BibitemShut {NoStop}%
\bibitem [{\citenamefont {Eremeev}\ \emph {et~al.}(2013)\citenamefont
  {Eremeev}, \citenamefont {Men'shov}, \citenamefont {Tugushev}, \citenamefont
  {Echenique},\ and\ \citenamefont {Chulkov}}]{eremeev2013prb}%
  \BibitemOpen
  \bibfield  {author} {\bibinfo {author} {\bibfnamefont {S.~V.}\ \bibnamefont
  {Eremeev}}, \bibinfo {author} {\bibfnamefont {V.~N.}\ \bibnamefont
  {Men'shov}}, \bibinfo {author} {\bibfnamefont {V.~V.}\ \bibnamefont
  {Tugushev}}, \bibinfo {author} {\bibfnamefont {P.~M.}\ \bibnamefont
  {Echenique}}, \ and\ \bibinfo {author} {\bibfnamefont {E.~V.}\ \bibnamefont
  {Chulkov}},\ }\href {\doibase 10.1103/PhysRevB.88.144430} {\bibfield
  {journal} {\bibinfo  {journal} {Phys. Rev. B}\ }\textbf {\bibinfo {volume}
  {88}},\ \bibinfo {pages} {144430} (\bibinfo {year} {2013})}\BibitemShut
  {NoStop}%
\bibitem [{\citenamefont {Luo}\ and\ \citenamefont {Qi}(2013)}]{luo2013prb}%
  \BibitemOpen
  \bibfield  {author} {\bibinfo {author} {\bibfnamefont {W.}~\bibnamefont
  {Luo}}\ and\ \bibinfo {author} {\bibfnamefont {X.-L.}\ \bibnamefont {Qi}},\
  }\href {\doibase 10.1103/PhysRevB.87.085431} {\bibfield  {journal} {\bibinfo
  {journal} {Phys. Rev. B}\ }\textbf {\bibinfo {volume} {87}},\ \bibinfo
  {pages} {085431} (\bibinfo {year} {2013})}\BibitemShut {NoStop}%
\bibitem [{\citenamefont {Hsu}\ \emph {et~al.}(2017)\citenamefont {Hsu},
  \citenamefont {Park},\ and\ \citenamefont {Kim}}]{hsu2017prb}%
  \BibitemOpen
  \bibfield  {author} {\bibinfo {author} {\bibfnamefont {Y.-T.}\ \bibnamefont
  {Hsu}}, \bibinfo {author} {\bibfnamefont {K.}~\bibnamefont {Park}}, \ and\
  \bibinfo {author} {\bibfnamefont {E.-A.}\ \bibnamefont {Kim}},\ }\href
  {\doibase 10.1103/PhysRevB.96.235433} {\bibfield  {journal} {\bibinfo
  {journal} {Phys. Rev. B}\ }\textbf {\bibinfo {volume} {96}},\ \bibinfo
  {pages} {235433} (\bibinfo {year} {2017})}\BibitemShut {NoStop}%
\bibitem [{\citenamefont {Spataru}\ and\ \citenamefont
  {L{\'e}onard}(2014)}]{spataru2014prb}%
  \BibitemOpen
  \bibfield  {author} {\bibinfo {author} {\bibfnamefont {C.~D.}\ \bibnamefont
  {Spataru}}\ and\ \bibinfo {author} {\bibfnamefont {F.}~\bibnamefont
  {L{\'e}onard}},\ }\href@noop {} {\bibfield  {journal} {\bibinfo  {journal}
  {Physical Review B}\ }\textbf {\bibinfo {volume} {90}},\ \bibinfo {pages}
  {085115} (\bibinfo {year} {2014})}\BibitemShut {NoStop}%
\bibitem [{\citenamefont {Yan}\ and\ \citenamefont {Felser}(2017)}]{yan2017ar}%
  \BibitemOpen
  \bibfield  {author} {\bibinfo {author} {\bibfnamefont {B.}~\bibnamefont
  {Yan}}\ and\ \bibinfo {author} {\bibfnamefont {C.}~\bibnamefont {Felser}},\
  }\href@noop {} {\bibfield  {journal} {\bibinfo  {journal} {Annual Review of
  Condensed Matter Physics}\ }\textbf {\bibinfo {volume} {8}},\ \bibinfo
  {pages} {337} (\bibinfo {year} {2017})}\BibitemShut {NoStop}%
\bibitem [{\citenamefont {Wan}\ \emph {et~al.}(2011)\citenamefont {Wan},
  \citenamefont {Turner}, \citenamefont {Vishwanath},\ and\ \citenamefont
  {Savrasov}}]{wan2011prb}%
  \BibitemOpen
  \bibfield  {author} {\bibinfo {author} {\bibfnamefont {X.}~\bibnamefont
  {Wan}}, \bibinfo {author} {\bibfnamefont {A.~M.}\ \bibnamefont {Turner}},
  \bibinfo {author} {\bibfnamefont {A.}~\bibnamefont {Vishwanath}}, \ and\
  \bibinfo {author} {\bibfnamefont {S.~Y.}\ \bibnamefont {Savrasov}},\
  }\href@noop {} {\bibfield  {journal} {\bibinfo  {journal} {Physical Review
  B—Condensed Matter and Materials Physics}\ }\textbf {\bibinfo {volume}
  {83}},\ \bibinfo {pages} {205101} (\bibinfo {year} {2011})}\BibitemShut
  {NoStop}%
\bibitem [{\citenamefont {Armitage}\ \emph {et~al.}(2018)\citenamefont
  {Armitage}, \citenamefont {Mele},\ and\ \citenamefont
  {Vishwanath}}]{armitageRMP2018}%
  \BibitemOpen
  \bibfield  {author} {\bibinfo {author} {\bibfnamefont {N.}~\bibnamefont
  {Armitage}}, \bibinfo {author} {\bibfnamefont {E.}~\bibnamefont {Mele}}, \
  and\ \bibinfo {author} {\bibfnamefont {A.}~\bibnamefont {Vishwanath}},\
  }\href@noop {} {\bibfield  {journal} {\bibinfo  {journal} {Reviews of Modern
  Physics}\ }\textbf {\bibinfo {volume} {90}},\ \bibinfo {pages} {015001}
  (\bibinfo {year} {2018})}\BibitemShut {NoStop}%
\bibitem [{\citenamefont {Sun}\ \emph {et~al.}(2016)\citenamefont {Sun},
  \citenamefont {Zhang}, \citenamefont {Felser},\ and\ \citenamefont
  {Yan}}]{sun2016prl}%
  \BibitemOpen
  \bibfield  {author} {\bibinfo {author} {\bibfnamefont {Y.}~\bibnamefont
  {Sun}}, \bibinfo {author} {\bibfnamefont {Y.}~\bibnamefont {Zhang}}, \bibinfo
  {author} {\bibfnamefont {C.}~\bibnamefont {Felser}}, \ and\ \bibinfo {author}
  {\bibfnamefont {B.}~\bibnamefont {Yan}},\ }\href@noop {} {\bibfield
  {journal} {\bibinfo  {journal} {Physical Review Letters}\ }\textbf {\bibinfo
  {volume} {117}},\ \bibinfo {pages} {146403} (\bibinfo {year}
  {2016})}\BibitemShut {NoStop}%
\bibitem [{\citenamefont {Han}\ \emph {et~al.}(2022)\citenamefont {Han},
  \citenamefont {Markou}, \citenamefont {Stensberg}, \citenamefont {Sun},
  \citenamefont {Felser},\ and\ \citenamefont {Wu}}]{han2022prb}%
  \BibitemOpen
  \bibfield  {author} {\bibinfo {author} {\bibfnamefont {X.}~\bibnamefont
  {Han}}, \bibinfo {author} {\bibfnamefont {A.}~\bibnamefont {Markou}},
  \bibinfo {author} {\bibfnamefont {J.}~\bibnamefont {Stensberg}}, \bibinfo
  {author} {\bibfnamefont {Y.}~\bibnamefont {Sun}}, \bibinfo {author}
  {\bibfnamefont {C.}~\bibnamefont {Felser}}, \ and\ \bibinfo {author}
  {\bibfnamefont {L.}~\bibnamefont {Wu}},\ }\href@noop {} {\bibfield  {journal}
  {\bibinfo  {journal} {Physical Review B}\ }\textbf {\bibinfo {volume}
  {105}},\ \bibinfo {pages} {174406} (\bibinfo {year} {2022})}\BibitemShut
  {NoStop}%
\bibitem [{\citenamefont {Manna}\ \emph {et~al.}(2018)\citenamefont {Manna},
  \citenamefont {Muechler}, \citenamefont {Kao}, \citenamefont {Stinshoff},
  \citenamefont {Zhang}, \citenamefont {Gooth}, \citenamefont {Kumar},
  \citenamefont {Kreiner}, \citenamefont {Koepernik}, \citenamefont {Car},
  \citenamefont {K\"ubler}, \citenamefont {Fecher}, \citenamefont {Shekhar},
  \citenamefont {Sun},\ and\ \citenamefont {Felser}}]{mannaPRX2018}%
  \BibitemOpen
  \bibfield  {author} {\bibinfo {author} {\bibfnamefont {K.}~\bibnamefont
  {Manna}}, \bibinfo {author} {\bibfnamefont {L.}~\bibnamefont {Muechler}},
  \bibinfo {author} {\bibfnamefont {T.-H.}\ \bibnamefont {Kao}}, \bibinfo
  {author} {\bibfnamefont {R.}~\bibnamefont {Stinshoff}}, \bibinfo {author}
  {\bibfnamefont {Y.}~\bibnamefont {Zhang}}, \bibinfo {author} {\bibfnamefont
  {J.}~\bibnamefont {Gooth}}, \bibinfo {author} {\bibfnamefont
  {N.}~\bibnamefont {Kumar}}, \bibinfo {author} {\bibfnamefont
  {G.}~\bibnamefont {Kreiner}}, \bibinfo {author} {\bibfnamefont
  {K.}~\bibnamefont {Koepernik}}, \bibinfo {author} {\bibfnamefont
  {R.}~\bibnamefont {Car}}, \bibinfo {author} {\bibfnamefont {J.}~\bibnamefont
  {K\"ubler}}, \bibinfo {author} {\bibfnamefont {G.~H.}\ \bibnamefont
  {Fecher}}, \bibinfo {author} {\bibfnamefont {C.}~\bibnamefont {Shekhar}},
  \bibinfo {author} {\bibfnamefont {Y.}~\bibnamefont {Sun}}, \ and\ \bibinfo
  {author} {\bibfnamefont {C.}~\bibnamefont {Felser}},\ }\href {\doibase
  10.1103/PhysRevX.8.041045} {\bibfield  {journal} {\bibinfo  {journal} {Phys.
  Rev. X}\ }\textbf {\bibinfo {volume} {8}},\ \bibinfo {pages} {041045}
  (\bibinfo {year} {2018})}\BibitemShut {NoStop}%
\bibitem [{\citenamefont {Liu}\ \emph {et~al.}(2018)\citenamefont {Liu},
  \citenamefont {Sun}, \citenamefont {Kumar}, \citenamefont {Muechler},
  \citenamefont {Sun}, \citenamefont {Jiao}, \citenamefont {Yang},
  \citenamefont {Liu}, \citenamefont {Liang}, \citenamefont {Xu} \emph
  {et~al.}}]{liu2018giant}%
  \BibitemOpen
  \bibfield  {author} {\bibinfo {author} {\bibfnamefont {E.}~\bibnamefont
  {Liu}}, \bibinfo {author} {\bibfnamefont {Y.}~\bibnamefont {Sun}}, \bibinfo
  {author} {\bibfnamefont {N.}~\bibnamefont {Kumar}}, \bibinfo {author}
  {\bibfnamefont {L.}~\bibnamefont {Muechler}}, \bibinfo {author}
  {\bibfnamefont {A.}~\bibnamefont {Sun}}, \bibinfo {author} {\bibfnamefont
  {L.}~\bibnamefont {Jiao}}, \bibinfo {author} {\bibfnamefont {S.-Y.}\
  \bibnamefont {Yang}}, \bibinfo {author} {\bibfnamefont {D.}~\bibnamefont
  {Liu}}, \bibinfo {author} {\bibfnamefont {A.}~\bibnamefont {Liang}}, \bibinfo
  {author} {\bibfnamefont {Q.}~\bibnamefont {Xu}},  \emph {et~al.},\
  }\href@noop {} {\bibfield  {journal} {\bibinfo  {journal} {Nature physics}\
  }\textbf {\bibinfo {volume} {14}},\ \bibinfo {pages} {1125} (\bibinfo {year}
  {2018})}\BibitemShut {NoStop}%
\bibitem [{\citenamefont {Huang}\ \emph {et~al.}(2008)\citenamefont {Huang},
  \citenamefont {Chen},\ and\ \citenamefont {Chien}}]{huang2008spin}%
  \BibitemOpen
  \bibfield  {author} {\bibinfo {author} {\bibfnamefont {S.}~\bibnamefont
  {Huang}}, \bibinfo {author} {\bibfnamefont {T.}~\bibnamefont {Chen}}, \ and\
  \bibinfo {author} {\bibfnamefont {C.}~\bibnamefont {Chien}},\ }\href@noop {}
  {\bibfield  {journal} {\bibinfo  {journal} {Applied Physics Letters}\
  }\textbf {\bibinfo {volume} {92}} (\bibinfo {year} {2008})}\BibitemShut
  {NoStop}%
\bibitem [{\citenamefont {Jen}\ \emph {et~al.}(2006{\natexlab{a}})\citenamefont
  {Jen}, \citenamefont {Yao}, \citenamefont {Chen}, \citenamefont {Wu},
  \citenamefont {Lee}, \citenamefont {Tsai},\ and\ \citenamefont
  {Chang}}]{jen2006jap}%
  \BibitemOpen
  \bibfield  {author} {\bibinfo {author} {\bibfnamefont {S.}~\bibnamefont
  {Jen}}, \bibinfo {author} {\bibfnamefont {Y.}~\bibnamefont {Yao}}, \bibinfo
  {author} {\bibfnamefont {Y.}~\bibnamefont {Chen}}, \bibinfo {author}
  {\bibfnamefont {J.}~\bibnamefont {Wu}}, \bibinfo {author} {\bibfnamefont
  {C.}~\bibnamefont {Lee}}, \bibinfo {author} {\bibfnamefont {T.}~\bibnamefont
  {Tsai}}, \ and\ \bibinfo {author} {\bibfnamefont {Y.}~\bibnamefont {Chang}},\
  }\href@noop {} {\bibfield  {journal} {\bibinfo  {journal} {Journal of applied
  physics}\ }\textbf {\bibinfo {volume} {99}} (\bibinfo {year}
  {2006}{\natexlab{a}})}\BibitemShut {NoStop}%
\bibitem [{\citenamefont {Bedoya-Pinto}\ \emph {et~al.}(2020)\citenamefont
  {Bedoya-Pinto}, \citenamefont {Pandeya}, \citenamefont {Liu}, \citenamefont
  {Deniz}, \citenamefont {Chang}, \citenamefont {Tan}, \citenamefont {Han},
  \citenamefont {Jena}, \citenamefont {Kostanovskiy},\ and\ \citenamefont
  {Parkin}}]{bedoyapinto2020acsnano}%
  \BibitemOpen
  \bibfield  {author} {\bibinfo {author} {\bibfnamefont {A.}~\bibnamefont
  {Bedoya-Pinto}}, \bibinfo {author} {\bibfnamefont {A.~K.}\ \bibnamefont
  {Pandeya}}, \bibinfo {author} {\bibfnamefont {D.}~\bibnamefont {Liu}},
  \bibinfo {author} {\bibfnamefont {H.}~\bibnamefont {Deniz}}, \bibinfo
  {author} {\bibfnamefont {K.}~\bibnamefont {Chang}}, \bibinfo {author}
  {\bibfnamefont {H.}~\bibnamefont {Tan}}, \bibinfo {author} {\bibfnamefont
  {H.}~\bibnamefont {Han}}, \bibinfo {author} {\bibfnamefont {J.}~\bibnamefont
  {Jena}}, \bibinfo {author} {\bibfnamefont {I.}~\bibnamefont {Kostanovskiy}},
  \ and\ \bibinfo {author} {\bibfnamefont {S.~S.}\ \bibnamefont {Parkin}},\
  }\href@noop {} {\bibfield  {journal} {\bibinfo  {journal} {ACS nano}\
  }\textbf {\bibinfo {volume} {14}},\ \bibinfo {pages} {4405} (\bibinfo {year}
  {2020})}\BibitemShut {NoStop}%
\bibitem [{\citenamefont {Han}\ \emph {et~al.}(2024)\citenamefont {Han},
  \citenamefont {Yi}, \citenamefont {Oh},\ and\ \citenamefont
  {Wu}}]{han2024nanolet}%
  \BibitemOpen
  \bibfield  {author} {\bibinfo {author} {\bibfnamefont {X.}~\bibnamefont
  {Han}}, \bibinfo {author} {\bibfnamefont {H.~T.}\ \bibnamefont {Yi}},
  \bibinfo {author} {\bibfnamefont {S.}~\bibnamefont {Oh}}, \ and\ \bibinfo
  {author} {\bibfnamefont {L.}~\bibnamefont {Wu}},\ }\href@noop {} {\bibfield
  {journal} {\bibinfo  {journal} {Nano Letters}\ }\textbf {\bibinfo {volume}
  {24}},\ \bibinfo {pages} {914} (\bibinfo {year} {2024})}\BibitemShut
  {NoStop}%
\bibitem [{\citenamefont {Stensberg}\ \emph {et~al.}(2023)\citenamefont
  {Stensberg}, \citenamefont {Han}, \citenamefont {Lee}, \citenamefont
  {McGill}, \citenamefont {Paglione}, \citenamefont {Takeuchi}, \citenamefont
  {Kane},\ and\ \citenamefont {Wu}}]{stensberg2023prl}%
  \BibitemOpen
  \bibfield  {author} {\bibinfo {author} {\bibfnamefont {J.}~\bibnamefont
  {Stensberg}}, \bibinfo {author} {\bibfnamefont {X.}~\bibnamefont {Han}},
  \bibinfo {author} {\bibfnamefont {S.}~\bibnamefont {Lee}}, \bibinfo {author}
  {\bibfnamefont {S.~A.}\ \bibnamefont {McGill}}, \bibinfo {author}
  {\bibfnamefont {J.}~\bibnamefont {Paglione}}, \bibinfo {author}
  {\bibfnamefont {I.}~\bibnamefont {Takeuchi}}, \bibinfo {author}
  {\bibfnamefont {C.~L.}\ \bibnamefont {Kane}}, \ and\ \bibinfo {author}
  {\bibfnamefont {L.}~\bibnamefont {Wu}},\ }\href@noop {} {\bibfield  {journal}
  {\bibinfo  {journal} {Physical Review Letters}\ }\textbf {\bibinfo {volume}
  {130}},\ \bibinfo {pages} {096901} (\bibinfo {year} {2023})}\BibitemShut
  {NoStop}%
\bibitem [{\citenamefont {Braun}\ \emph {et~al.}(2016)\citenamefont {Braun},
  \citenamefont {Mussler}, \citenamefont {Hruban}, \citenamefont
  {Konczykowski}, \citenamefont {Schumann}, \citenamefont {Wolf}, \citenamefont
  {M{\"u}nzenberg}, \citenamefont {Perfetti},\ and\ \citenamefont
  {Kampfrath}}]{braun2016ulNatCom}%
  \BibitemOpen
  \bibfield  {author} {\bibinfo {author} {\bibfnamefont {L.}~\bibnamefont
  {Braun}}, \bibinfo {author} {\bibfnamefont {G.}~\bibnamefont {Mussler}},
  \bibinfo {author} {\bibfnamefont {A.}~\bibnamefont {Hruban}}, \bibinfo
  {author} {\bibfnamefont {M.}~\bibnamefont {Konczykowski}}, \bibinfo {author}
  {\bibfnamefont {T.}~\bibnamefont {Schumann}}, \bibinfo {author}
  {\bibfnamefont {M.}~\bibnamefont {Wolf}}, \bibinfo {author} {\bibfnamefont
  {M.}~\bibnamefont {M{\"u}nzenberg}}, \bibinfo {author} {\bibfnamefont
  {L.}~\bibnamefont {Perfetti}}, \ and\ \bibinfo {author} {\bibfnamefont
  {T.}~\bibnamefont {Kampfrath}},\ }\href@noop {} {\bibfield  {journal}
  {\bibinfo  {journal} {Nature communications}\ }\textbf {\bibinfo {volume}
  {7}},\ \bibinfo {pages} {1} (\bibinfo {year} {2016})}\BibitemShut {NoStop}%
\bibitem [{\citenamefont {Stensberg}\ \emph {et~al.}(2024)\citenamefont
  {Stensberg}, \citenamefont {Han}, \citenamefont {Ni}, \citenamefont {Yao},
  \citenamefont {Yuan}, \citenamefont {Mallick}, \citenamefont {Gandhi},
  \citenamefont {Oh},\ and\ \citenamefont {Wu}}]{stensberg2024prb}%
  \BibitemOpen
  \bibfield  {author} {\bibinfo {author} {\bibfnamefont {J.}~\bibnamefont
  {Stensberg}}, \bibinfo {author} {\bibfnamefont {X.}~\bibnamefont {Han}},
  \bibinfo {author} {\bibfnamefont {Z.}~\bibnamefont {Ni}}, \bibinfo {author}
  {\bibfnamefont {X.}~\bibnamefont {Yao}}, \bibinfo {author} {\bibfnamefont
  {X.}~\bibnamefont {Yuan}}, \bibinfo {author} {\bibfnamefont {D.}~\bibnamefont
  {Mallick}}, \bibinfo {author} {\bibfnamefont {A.}~\bibnamefont {Gandhi}},
  \bibinfo {author} {\bibfnamefont {S.}~\bibnamefont {Oh}}, \ and\ \bibinfo
  {author} {\bibfnamefont {L.}~\bibnamefont {Wu}},\ }\href {\doibase
  10.1103/PhysRevB.109.245112} {\bibfield  {journal} {\bibinfo  {journal}
  {Phys. Rev. B}\ }\textbf {\bibinfo {volume} {109}},\ \bibinfo {pages}
  {245112} (\bibinfo {year} {2024})}\BibitemShut {NoStop}%
\bibitem [{\citenamefont {Jen}\ \emph {et~al.}(2006{\natexlab{b}})\citenamefont
  {Jen}, \citenamefont {Yao}, \citenamefont {Chen}, \citenamefont {Wu},
  \citenamefont {Lee}, \citenamefont {Tsai},\ and\ \citenamefont
  {Chang}}]{jen2006magnetic}%
  \BibitemOpen
  \bibfield  {author} {\bibinfo {author} {\bibfnamefont {S.}~\bibnamefont
  {Jen}}, \bibinfo {author} {\bibfnamefont {Y.}~\bibnamefont {Yao}}, \bibinfo
  {author} {\bibfnamefont {Y.}~\bibnamefont {Chen}}, \bibinfo {author}
  {\bibfnamefont {J.}~\bibnamefont {Wu}}, \bibinfo {author} {\bibfnamefont
  {C.}~\bibnamefont {Lee}}, \bibinfo {author} {\bibfnamefont {T.}~\bibnamefont
  {Tsai}}, \ and\ \bibinfo {author} {\bibfnamefont {Y.}~\bibnamefont {Chang}},\
  }\href@noop {} {\bibfield  {journal} {\bibinfo  {journal} {Journal of applied
  physics}\ }\textbf {\bibinfo {volume} {99}} (\bibinfo {year}
  {2006}{\natexlab{b}})}\BibitemShut {NoStop}%
\bibitem [{\citenamefont {Wang}\ \emph {et~al.}(2012)\citenamefont {Wang},
  \citenamefont {Bian}, \citenamefont {Miller},\ and\ \citenamefont
  {Chiang}}]{wang2012prl}%
  \BibitemOpen
  \bibfield  {author} {\bibinfo {author} {\bibfnamefont {X.}~\bibnamefont
  {Wang}}, \bibinfo {author} {\bibfnamefont {G.}~\bibnamefont {Bian}}, \bibinfo
  {author} {\bibfnamefont {T.}~\bibnamefont {Miller}}, \ and\ \bibinfo {author}
  {\bibfnamefont {T.-C.}\ \bibnamefont {Chiang}},\ }\href {\doibase
  10.1103/PhysRevLett.108.096404} {\bibfield  {journal} {\bibinfo  {journal}
  {Phys. Rev. Lett.}\ }\textbf {\bibinfo {volume} {108}},\ \bibinfo {pages}
  {096404} (\bibinfo {year} {2012})}\BibitemShut {NoStop}%
\bibitem [{\citenamefont {Yashina}\ \emph {et~al.}(2013)\citenamefont
  {Yashina}, \citenamefont {S{\'a}nchez-Barriga}, \citenamefont {Scholz},
  \citenamefont {Volykhov}, \citenamefont {Sirotina}, \citenamefont
  {Neudachina}, \citenamefont {Tamm}, \citenamefont {Varykhalov}, \citenamefont
  {Marchenko}, \citenamefont {Springholz} \emph {et~al.}}]{yashina2013acsnano}%
  \BibitemOpen
  \bibfield  {author} {\bibinfo {author} {\bibfnamefont {L.~V.}\ \bibnamefont
  {Yashina}}, \bibinfo {author} {\bibfnamefont {J.}~\bibnamefont
  {S{\'a}nchez-Barriga}}, \bibinfo {author} {\bibfnamefont {M.~R.}\
  \bibnamefont {Scholz}}, \bibinfo {author} {\bibfnamefont {A.~A.}\
  \bibnamefont {Volykhov}}, \bibinfo {author} {\bibfnamefont {A.~P.}\
  \bibnamefont {Sirotina}}, \bibinfo {author} {\bibfnamefont {S.}~\bibnamefont
  {Neudachina}, \bibfnamefont {Vera}}, \bibinfo {author} {\bibfnamefont
  {M.~E.}\ \bibnamefont {Tamm}}, \bibinfo {author} {\bibfnamefont
  {A.}~\bibnamefont {Varykhalov}}, \bibinfo {author} {\bibfnamefont
  {D.}~\bibnamefont {Marchenko}}, \bibinfo {author} {\bibfnamefont
  {G.}~\bibnamefont {Springholz}},  \emph {et~al.},\ }\href@noop {} {\bibfield
  {journal} {\bibinfo  {journal} {Acs Nano}\ }\textbf {\bibinfo {volume} {7}},\
  \bibinfo {pages} {5181} (\bibinfo {year} {2013})}\BibitemShut {NoStop}%
\bibitem [{\citenamefont {Yang}\ \emph {et~al.}(2020)\citenamefont {Yang},
  \citenamefont {Zheng}, \citenamefont {Chen}, \citenamefont {Xu},
  \citenamefont {Wang},\ and\ \citenamefont {Xu}}]{yang2020jpcc}%
  \BibitemOpen
  \bibfield  {author} {\bibinfo {author} {\bibfnamefont {J.}~\bibnamefont
  {Yang}}, \bibinfo {author} {\bibfnamefont {B.}~\bibnamefont {Zheng}},
  \bibinfo {author} {\bibfnamefont {Z.}~\bibnamefont {Chen}}, \bibinfo {author}
  {\bibfnamefont {W.}~\bibnamefont {Xu}}, \bibinfo {author} {\bibfnamefont
  {R.}~\bibnamefont {Wang}}, \ and\ \bibinfo {author} {\bibfnamefont
  {H.}~\bibnamefont {Xu}},\ }\href@noop {} {\bibfield  {journal} {\bibinfo
  {journal} {The Journal of Physical Chemistry C}\ }\textbf {\bibinfo {volume}
  {124}},\ \bibinfo {pages} {6253} (\bibinfo {year} {2020})}\BibitemShut
  {NoStop}%
\bibitem [{\citenamefont {Green}\ \emph {et~al.}(2016)\citenamefont {Green},
  \citenamefont {Dey}, \citenamefont {An}, \citenamefont {O'Brien},
  \citenamefont {O'Mullane}, \citenamefont {Thiel},\ and\ \citenamefont
  {Diebold}}]{green2016jvst}%
  \BibitemOpen
  \bibfield  {author} {\bibinfo {author} {\bibfnamefont {A.~J.}\ \bibnamefont
  {Green}}, \bibinfo {author} {\bibfnamefont {S.}~\bibnamefont {Dey}}, \bibinfo
  {author} {\bibfnamefont {Y.~Q.}\ \bibnamefont {An}}, \bibinfo {author}
  {\bibfnamefont {B.}~\bibnamefont {O'Brien}}, \bibinfo {author} {\bibfnamefont
  {S.}~\bibnamefont {O'Mullane}}, \bibinfo {author} {\bibfnamefont
  {B.}~\bibnamefont {Thiel}}, \ and\ \bibinfo {author} {\bibfnamefont {A.~C.}\
  \bibnamefont {Diebold}},\ }\href@noop {} {\bibfield  {journal} {\bibinfo
  {journal} {Journal of Vacuum Science \& Technology A}\ }\textbf {\bibinfo
  {volume} {34}} (\bibinfo {year} {2016})}\BibitemShut {NoStop}%
\bibitem [{\citenamefont {He}\ \emph {et~al.}(2015)\citenamefont {He},
  \citenamefont {Li}, \citenamefont {Chen},\ and\ \citenamefont
  {Wu}}]{he2015scirep}%
  \BibitemOpen
  \bibfield  {author} {\bibinfo {author} {\bibfnamefont {X.}~\bibnamefont
  {He}}, \bibinfo {author} {\bibfnamefont {H.}~\bibnamefont {Li}}, \bibinfo
  {author} {\bibfnamefont {L.}~\bibnamefont {Chen}}, \ and\ \bibinfo {author}
  {\bibfnamefont {K.}~\bibnamefont {Wu}},\ }\href@noop {} {\bibfield  {journal}
  {\bibinfo  {journal} {Scientific reports}\ }\textbf {\bibinfo {volume} {5}},\
  \bibinfo {pages} {8830} (\bibinfo {year} {2015})}\BibitemShut {NoStop}%
\bibitem [{\citenamefont {Kong}\ \emph {et~al.}(2011)\citenamefont {Kong},
  \citenamefont {Chen}, \citenamefont {Cha}, \citenamefont {Zhang},
  \citenamefont {Analytis}, \citenamefont {Lai}, \citenamefont {Liu},
  \citenamefont {Hong}, \citenamefont {Koski}, \citenamefont {Mo} \emph
  {et~al.}}]{kong2011natnano}%
  \BibitemOpen
  \bibfield  {author} {\bibinfo {author} {\bibfnamefont {D.}~\bibnamefont
  {Kong}}, \bibinfo {author} {\bibfnamefont {Y.}~\bibnamefont {Chen}}, \bibinfo
  {author} {\bibfnamefont {J.~J.}\ \bibnamefont {Cha}}, \bibinfo {author}
  {\bibfnamefont {Q.}~\bibnamefont {Zhang}}, \bibinfo {author} {\bibfnamefont
  {J.~G.}\ \bibnamefont {Analytis}}, \bibinfo {author} {\bibfnamefont
  {K.}~\bibnamefont {Lai}}, \bibinfo {author} {\bibfnamefont {Z.}~\bibnamefont
  {Liu}}, \bibinfo {author} {\bibfnamefont {S.~S.}\ \bibnamefont {Hong}},
  \bibinfo {author} {\bibfnamefont {K.~J.}\ \bibnamefont {Koski}}, \bibinfo
  {author} {\bibfnamefont {S.-K.}\ \bibnamefont {Mo}},  \emph {et~al.},\
  }\href@noop {} {\bibfield  {journal} {\bibinfo  {journal} {Nature
  nanotechnology}\ }\textbf {\bibinfo {volume} {6}},\ \bibinfo {pages} {705}
  (\bibinfo {year} {2011})}\BibitemShut {NoStop}%
\bibitem [{\citenamefont {Zhang}\ \emph {et~al.}(2011)\citenamefont {Zhang},
  \citenamefont {Chang}, \citenamefont {Zhang}, \citenamefont {Wen},
  \citenamefont {Feng}, \citenamefont {Li}, \citenamefont {Liu}, \citenamefont
  {He}, \citenamefont {Wang}, \citenamefont {Chen} \emph
  {et~al.}}]{zhang_band_2011}%
  \BibitemOpen
  \bibfield  {author} {\bibinfo {author} {\bibfnamefont {J.}~\bibnamefont
  {Zhang}}, \bibinfo {author} {\bibfnamefont {C.-Z.}\ \bibnamefont {Chang}},
  \bibinfo {author} {\bibfnamefont {Z.}~\bibnamefont {Zhang}}, \bibinfo
  {author} {\bibfnamefont {J.}~\bibnamefont {Wen}}, \bibinfo {author}
  {\bibfnamefont {X.}~\bibnamefont {Feng}}, \bibinfo {author} {\bibfnamefont
  {K.}~\bibnamefont {Li}}, \bibinfo {author} {\bibfnamefont {M.}~\bibnamefont
  {Liu}}, \bibinfo {author} {\bibfnamefont {K.}~\bibnamefont {He}}, \bibinfo
  {author} {\bibfnamefont {L.}~\bibnamefont {Wang}}, \bibinfo {author}
  {\bibfnamefont {X.}~\bibnamefont {Chen}},  \emph {et~al.},\ }\href@noop {}
  {\bibfield  {journal} {\bibinfo  {journal} {Nature communications}\ }\textbf
  {\bibinfo {volume} {2}},\ \bibinfo {pages} {574} (\bibinfo {year}
  {2011})}\BibitemShut {NoStop}%
\bibitem [{\citenamefont {He}\ \emph {et~al.}(2012{\natexlab{a}})\citenamefont
  {He}, \citenamefont {Guan}, \citenamefont {Wang}, \citenamefont {Feng},
  \citenamefont {Cheng}, \citenamefont {Chen}, \citenamefont {Li},\ and\
  \citenamefont {Wu}}]{he_highly_2012}%
  \BibitemOpen
  \bibfield  {author} {\bibinfo {author} {\bibfnamefont {X.}~\bibnamefont
  {He}}, \bibinfo {author} {\bibfnamefont {T.}~\bibnamefont {Guan}}, \bibinfo
  {author} {\bibfnamefont {X.}~\bibnamefont {Wang}}, \bibinfo {author}
  {\bibfnamefont {B.}~\bibnamefont {Feng}}, \bibinfo {author} {\bibfnamefont
  {P.}~\bibnamefont {Cheng}}, \bibinfo {author} {\bibfnamefont
  {L.}~\bibnamefont {Chen}}, \bibinfo {author} {\bibfnamefont {Y.}~\bibnamefont
  {Li}}, \ and\ \bibinfo {author} {\bibfnamefont {K.}~\bibnamefont {Wu}},\
  }\href@noop {} {\bibfield  {journal} {\bibinfo  {journal} {Applied Physics
  Letters}\ }\textbf {\bibinfo {volume} {101}} (\bibinfo {year}
  {2012}{\natexlab{a}})}\BibitemShut {NoStop}%
\bibitem [{\citenamefont {He}\ \emph {et~al.}(2012{\natexlab{b}})\citenamefont
  {He}, \citenamefont {Guan}, \citenamefont {Wang}, \citenamefont {Feng},
  \citenamefont {Cheng}, \citenamefont {Chen}, \citenamefont {Li},\ and\
  \citenamefont {Wu}}]{niu_realization_2012}%
  \BibitemOpen
  \bibfield  {author} {\bibinfo {author} {\bibfnamefont {X.}~\bibnamefont
  {He}}, \bibinfo {author} {\bibfnamefont {T.}~\bibnamefont {Guan}}, \bibinfo
  {author} {\bibfnamefont {X.}~\bibnamefont {Wang}}, \bibinfo {author}
  {\bibfnamefont {B.}~\bibnamefont {Feng}}, \bibinfo {author} {\bibfnamefont
  {P.}~\bibnamefont {Cheng}}, \bibinfo {author} {\bibfnamefont
  {L.}~\bibnamefont {Chen}}, \bibinfo {author} {\bibfnamefont {Y.}~\bibnamefont
  {Li}}, \ and\ \bibinfo {author} {\bibfnamefont {K.}~\bibnamefont {Wu}},\
  }\href@noop {} {\bibfield  {journal} {\bibinfo  {journal} {Applied Physics
  Letters}\ }\textbf {\bibinfo {volume} {101}} (\bibinfo {year}
  {2012}{\natexlab{b}})}\BibitemShut {NoStop}%
\bibitem [{\citenamefont {Fu}(2009)}]{Fu2009PRL}%
  \BibitemOpen
  \bibfield  {author} {\bibinfo {author} {\bibfnamefont {L.}~\bibnamefont
  {Fu}},\ }\href@noop {} {\bibfield  {journal} {\bibinfo  {journal} {Physical
  review letters}\ }\textbf {\bibinfo {volume} {103}},\ \bibinfo {pages}
  {266801} (\bibinfo {year} {2009})}\BibitemShut {NoStop}%
\bibitem [{\citenamefont {Chen}\ \emph {et~al.}(2009)\citenamefont {Chen},
  \citenamefont {Analytis}, \citenamefont {Chu}, \citenamefont {Liu},
  \citenamefont {Mo}, \citenamefont {Qi}, \citenamefont {Zhang}, \citenamefont
  {Lu}, \citenamefont {Dai}, \citenamefont {Fang} \emph
  {et~al.}}]{chen_experimental_2009}%
  \BibitemOpen
  \bibfield  {author} {\bibinfo {author} {\bibfnamefont {Y.}~\bibnamefont
  {Chen}}, \bibinfo {author} {\bibfnamefont {J.~G.}\ \bibnamefont {Analytis}},
  \bibinfo {author} {\bibfnamefont {J.-H.}\ \bibnamefont {Chu}}, \bibinfo
  {author} {\bibfnamefont {Z.}~\bibnamefont {Liu}}, \bibinfo {author}
  {\bibfnamefont {S.-K.}\ \bibnamefont {Mo}}, \bibinfo {author} {\bibfnamefont
  {X.-L.}\ \bibnamefont {Qi}}, \bibinfo {author} {\bibfnamefont
  {H.}~\bibnamefont {Zhang}}, \bibinfo {author} {\bibfnamefont
  {D.}~\bibnamefont {Lu}}, \bibinfo {author} {\bibfnamefont {X.}~\bibnamefont
  {Dai}}, \bibinfo {author} {\bibfnamefont {Z.}~\bibnamefont {Fang}},  \emph
  {et~al.},\ }\href@noop {} {\bibfield  {journal} {\bibinfo  {journal}
  {science}\ }\textbf {\bibinfo {volume} {325}},\ \bibinfo {pages} {178}
  (\bibinfo {year} {2009})}\BibitemShut {NoStop}%
\bibitem [{\citenamefont {Bianchi}\ \emph {et~al.}(2010)\citenamefont
  {Bianchi}, \citenamefont {Guan}, \citenamefont {Bao}, \citenamefont {Mi},
  \citenamefont {Iversen}, \citenamefont {King},\ and\ \citenamefont
  {Hofmann}}]{Bianchi10}%
  \BibitemOpen
  \bibfield  {author} {\bibinfo {author} {\bibfnamefont {M.}~\bibnamefont
  {Bianchi}}, \bibinfo {author} {\bibfnamefont {D.}~\bibnamefont {Guan}},
  \bibinfo {author} {\bibfnamefont {S.}~\bibnamefont {Bao}}, \bibinfo {author}
  {\bibfnamefont {J.}~\bibnamefont {Mi}}, \bibinfo {author} {\bibfnamefont
  {B.~B.}\ \bibnamefont {Iversen}}, \bibinfo {author} {\bibfnamefont {P.~D.}\
  \bibnamefont {King}}, \ and\ \bibinfo {author} {\bibfnamefont
  {P.}~\bibnamefont {Hofmann}},\ }\href@noop {} {\bibfield  {journal} {\bibinfo
   {journal} {Nature communications}\ }\textbf {\bibinfo {volume} {1}},\
  \bibinfo {pages} {128} (\bibinfo {year} {2010})}\BibitemShut {NoStop}%
\bibitem [{\citenamefont {Pauly}\ \emph {et~al.}(2012)\citenamefont {Pauly},
  \citenamefont {Bihlmayer}, \citenamefont {Liebmann}, \citenamefont {Grob},
  \citenamefont {Georgi}, \citenamefont {Subramaniam}, \citenamefont {Scholz},
  \citenamefont {S{\'a}nchez-Barriga}, \citenamefont {Varykhalov},
  \citenamefont {Bl{\"u}gel} \emph {et~al.}}]{pauly2012prb}%
  \BibitemOpen
  \bibfield  {author} {\bibinfo {author} {\bibfnamefont {C.}~\bibnamefont
  {Pauly}}, \bibinfo {author} {\bibfnamefont {G.}~\bibnamefont {Bihlmayer}},
  \bibinfo {author} {\bibfnamefont {M.}~\bibnamefont {Liebmann}}, \bibinfo
  {author} {\bibfnamefont {M.}~\bibnamefont {Grob}}, \bibinfo {author}
  {\bibfnamefont {A.}~\bibnamefont {Georgi}}, \bibinfo {author} {\bibfnamefont
  {D.}~\bibnamefont {Subramaniam}}, \bibinfo {author} {\bibfnamefont
  {M.}~\bibnamefont {Scholz}}, \bibinfo {author} {\bibfnamefont
  {J.}~\bibnamefont {S{\'a}nchez-Barriga}}, \bibinfo {author} {\bibfnamefont
  {A.}~\bibnamefont {Varykhalov}}, \bibinfo {author} {\bibfnamefont
  {S.}~\bibnamefont {Bl{\"u}gel}},  \emph {et~al.},\ }\href@noop {} {\bibfield
  {journal} {\bibinfo  {journal} {Physical Review B—Condensed Matter and
  Materials Physics}\ }\textbf {\bibinfo {volume} {86}},\ \bibinfo {pages}
  {235106} (\bibinfo {year} {2012})}\BibitemShut {NoStop}%
\bibitem [{\citenamefont {Zhu}\ \emph {et~al.}(2011)\citenamefont {Zhu},
  \citenamefont {Levy}, \citenamefont {Ludbrook}, \citenamefont {Veenstra},
  \citenamefont {Rosen}, \citenamefont {Comin}, \citenamefont {Wong},
  \citenamefont {Dosanjh}, \citenamefont {Ubaldini}, \citenamefont {Syers}
  \emph {et~al.}}]{zhu2011prl}%
  \BibitemOpen
  \bibfield  {author} {\bibinfo {author} {\bibfnamefont {Z.-H.}\ \bibnamefont
  {Zhu}}, \bibinfo {author} {\bibfnamefont {G.}~\bibnamefont {Levy}}, \bibinfo
  {author} {\bibfnamefont {B.}~\bibnamefont {Ludbrook}}, \bibinfo {author}
  {\bibfnamefont {C.}~\bibnamefont {Veenstra}}, \bibinfo {author}
  {\bibfnamefont {J.}~\bibnamefont {Rosen}}, \bibinfo {author} {\bibfnamefont
  {R.}~\bibnamefont {Comin}}, \bibinfo {author} {\bibfnamefont
  {D.}~\bibnamefont {Wong}}, \bibinfo {author} {\bibfnamefont {P.}~\bibnamefont
  {Dosanjh}}, \bibinfo {author} {\bibfnamefont {A.}~\bibnamefont {Ubaldini}},
  \bibinfo {author} {\bibfnamefont {P.}~\bibnamefont {Syers}},  \emph
  {et~al.},\ }\href@noop {} {\bibfield  {journal} {\bibinfo  {journal}
  {Physical review letters}\ }\textbf {\bibinfo {volume} {107}},\ \bibinfo
  {pages} {186405} (\bibinfo {year} {2011})}\BibitemShut {NoStop}%
\bibitem [{\citenamefont {Wang}\ \emph {et~al.}(2015)\citenamefont {Wang},
  \citenamefont {Tang}, \citenamefont {Wan}, \citenamefont {Fedorov},
  \citenamefont {Miotkowski}, \citenamefont {Chen}, \citenamefont {Duan},\ and\
  \citenamefont {Zhou}}]{wang2015nanoletters}%
  \BibitemOpen
  \bibfield  {author} {\bibinfo {author} {\bibfnamefont {E.}~\bibnamefont
  {Wang}}, \bibinfo {author} {\bibfnamefont {P.}~\bibnamefont {Tang}}, \bibinfo
  {author} {\bibfnamefont {G.}~\bibnamefont {Wan}}, \bibinfo {author}
  {\bibfnamefont {A.~V.}\ \bibnamefont {Fedorov}}, \bibinfo {author}
  {\bibfnamefont {I.}~\bibnamefont {Miotkowski}}, \bibinfo {author}
  {\bibfnamefont {Y.~P.}\ \bibnamefont {Chen}}, \bibinfo {author}
  {\bibfnamefont {W.}~\bibnamefont {Duan}}, \ and\ \bibinfo {author}
  {\bibfnamefont {S.}~\bibnamefont {Zhou}},\ }\href@noop {} {\bibfield
  {journal} {\bibinfo  {journal} {Nano letters}\ }\textbf {\bibinfo {volume}
  {15}},\ \bibinfo {pages} {2031} (\bibinfo {year} {2015})}\BibitemShut
  {NoStop}%
\bibitem [{\citenamefont {Souma}\ \emph {et~al.}(2016)\citenamefont {Souma},
  \citenamefont {Wang}, \citenamefont {Kotaka}, \citenamefont {Sato},
  \citenamefont {Nakayama}, \citenamefont {Tanaka}, \citenamefont {Kimizuka},
  \citenamefont {Takahashi}, \citenamefont {Yamauchi}, \citenamefont {Oguchi},
  \citenamefont {Segawa},\ and\ \citenamefont {Ando}}]{souma2016prb}%
  \BibitemOpen
  \bibfield  {author} {\bibinfo {author} {\bibfnamefont {S.}~\bibnamefont
  {Souma}}, \bibinfo {author} {\bibfnamefont {Z.}~\bibnamefont {Wang}},
  \bibinfo {author} {\bibfnamefont {H.}~\bibnamefont {Kotaka}}, \bibinfo
  {author} {\bibfnamefont {T.}~\bibnamefont {Sato}}, \bibinfo {author}
  {\bibfnamefont {K.}~\bibnamefont {Nakayama}}, \bibinfo {author}
  {\bibfnamefont {Y.}~\bibnamefont {Tanaka}}, \bibinfo {author} {\bibfnamefont
  {H.}~\bibnamefont {Kimizuka}}, \bibinfo {author} {\bibfnamefont
  {T.}~\bibnamefont {Takahashi}}, \bibinfo {author} {\bibfnamefont
  {K.}~\bibnamefont {Yamauchi}}, \bibinfo {author} {\bibfnamefont
  {T.}~\bibnamefont {Oguchi}}, \bibinfo {author} {\bibfnamefont
  {K.}~\bibnamefont {Segawa}}, \ and\ \bibinfo {author} {\bibfnamefont
  {Y.}~\bibnamefont {Ando}},\ }\href {\doibase 10.1103/PhysRevB.93.161112}
  {\bibfield  {journal} {\bibinfo  {journal} {Phys. Rev. B}\ }\textbf {\bibinfo
  {volume} {93}},\ \bibinfo {pages} {161112} (\bibinfo {year}
  {2016})}\BibitemShut {NoStop}%
\bibitem [{\citenamefont {Liu}\ \emph {et~al.}(2016)\citenamefont {Liu},
  \citenamefont {Yang}, \citenamefont {Sun}, \citenamefont {Zhang},
  \citenamefont {Peng}, \citenamefont {Yang}, \citenamefont {Chen},
  \citenamefont {Zhang}, \citenamefont {Guo}, \citenamefont {Prabhakaran} \emph
  {et~al.}}]{liu2016evolution}%
  \BibitemOpen
  \bibfield  {author} {\bibinfo {author} {\bibfnamefont {Z.}~\bibnamefont
  {Liu}}, \bibinfo {author} {\bibfnamefont {L.}~\bibnamefont {Yang}}, \bibinfo
  {author} {\bibfnamefont {Y.}~\bibnamefont {Sun}}, \bibinfo {author}
  {\bibfnamefont {T.}~\bibnamefont {Zhang}}, \bibinfo {author} {\bibfnamefont
  {H.}~\bibnamefont {Peng}}, \bibinfo {author} {\bibfnamefont {H.}~\bibnamefont
  {Yang}}, \bibinfo {author} {\bibfnamefont {C.}~\bibnamefont {Chen}}, \bibinfo
  {author} {\bibfnamefont {Y.~f.}\ \bibnamefont {Zhang}}, \bibinfo {author}
  {\bibfnamefont {Y.}~\bibnamefont {Guo}}, \bibinfo {author} {\bibfnamefont
  {D.}~\bibnamefont {Prabhakaran}},  \emph {et~al.},\ }\href@noop {} {\bibfield
   {journal} {\bibinfo  {journal} {Nature materials}\ }\textbf {\bibinfo
  {volume} {15}},\ \bibinfo {pages} {27} (\bibinfo {year} {2016})}\BibitemShut
  {NoStop}%
\bibitem [{\citenamefont {Bedoya-Pinto}\ \emph {et~al.}(2021)\citenamefont
  {Bedoya-Pinto}, \citenamefont {Liu}, \citenamefont {Tan}, \citenamefont
  {Pandeya}, \citenamefont {Chang}, \citenamefont {Zhang},\ and\ \citenamefont
  {Parkin}}]{bedoya2021advmat}%
  \BibitemOpen
  \bibfield  {author} {\bibinfo {author} {\bibfnamefont {A.}~\bibnamefont
  {Bedoya-Pinto}}, \bibinfo {author} {\bibfnamefont {D.}~\bibnamefont {Liu}},
  \bibinfo {author} {\bibfnamefont {H.}~\bibnamefont {Tan}}, \bibinfo {author}
  {\bibfnamefont {A.~K.}\ \bibnamefont {Pandeya}}, \bibinfo {author}
  {\bibfnamefont {K.}~\bibnamefont {Chang}}, \bibinfo {author} {\bibfnamefont
  {J.}~\bibnamefont {Zhang}}, \ and\ \bibinfo {author} {\bibfnamefont {S.~S.}\
  \bibnamefont {Parkin}},\ }\href@noop {} {\bibfield  {journal} {\bibinfo
  {journal} {Advanced Materials}\ }\textbf {\bibinfo {volume} {33}},\ \bibinfo
  {pages} {2008634} (\bibinfo {year} {2021})}\BibitemShut {NoStop}%
\bibitem [{\citenamefont {Kampfrath}\ \emph {et~al.}(2007)\citenamefont
  {Kampfrath}, \citenamefont {N{\"o}tzold},\ and\ \citenamefont
  {Wolf}}]{kampfrath2007sampling}%
  \BibitemOpen
  \bibfield  {author} {\bibinfo {author} {\bibfnamefont {T.}~\bibnamefont
  {Kampfrath}}, \bibinfo {author} {\bibfnamefont {J.}~\bibnamefont
  {N{\"o}tzold}}, \ and\ \bibinfo {author} {\bibfnamefont {M.}~\bibnamefont
  {Wolf}},\ }\href@noop {} {\bibfield  {journal} {\bibinfo  {journal} {Applied
  physics letters}\ }\textbf {\bibinfo {volume} {90}} (\bibinfo {year}
  {2007})}\BibitemShut {NoStop}%
\bibitem [{\citenamefont {Leitenstorfer}\ \emph {et~al.}(1999)\citenamefont
  {Leitenstorfer}, \citenamefont {Hunsche}, \citenamefont {Shah}, \citenamefont
  {Nuss},\ and\ \citenamefont {Knox}}]{leitenstorfer1999detectors}%
  \BibitemOpen
  \bibfield  {author} {\bibinfo {author} {\bibfnamefont {A.}~\bibnamefont
  {Leitenstorfer}}, \bibinfo {author} {\bibfnamefont {S.}~\bibnamefont
  {Hunsche}}, \bibinfo {author} {\bibfnamefont {J.}~\bibnamefont {Shah}},
  \bibinfo {author} {\bibfnamefont {M.}~\bibnamefont {Nuss}}, \ and\ \bibinfo
  {author} {\bibfnamefont {W.}~\bibnamefont {Knox}},\ }\href@noop {} {\bibfield
   {journal} {\bibinfo  {journal} {Applied physics letters}\ }\textbf {\bibinfo
  {volume} {74}},\ \bibinfo {pages} {1516} (\bibinfo {year}
  {1999})}\BibitemShut {NoStop}%
\end{thebibliography}%

\end{document}